\documentclass[aps,twocolumn,groupedaddress,nofootinbib]{revtex4}

\usepackage[pdftex]{graphicx}
\usepackage{hyperref}
\usepackage{color}
\usepackage[english]{babel}
\usepackage{mhchem}

\usepackage{pifont}% http://ctan.org/pkg/pifont
\newcommand{\cmark}{\ding{51}}%
\newcommand{\xmark}{\ding{55}}%

\newcommand{\be}{\begin{equation}}
\newcommand{\ee}{\end{equation}}

\begin{document}

\title{Leveraging percolation theory to single out influential spreaders in networks}

\author{Filippo Radicchi}
\affiliation{Center for Complex Networks and Systems Research, School of Informatics and Computing, Indiana University, Bloomington, USA}
\email{filiradi@indiana.edu}

\author{Claudio Castellano}
\affiliation{Istituto dei Sistemi Complessi (ISC-CNR), Via dei Taurini 19, 00185 Roma, Italy,\\ and
Dipartimento di Fisica, Sapienza Universit\`a di Roma, Roma, Italy}

%\date{\today}

\begin{abstract}
Among the consequences of the disordered interaction topology 
underlying many social, technological and biological systems, a particularly
important one is that some nodes, just because
of their position in the network, may have a disproportionate effect on
dynamical processes mediated by the complex interaction pattern.
For example, the  early adoption by an opinion leader 
in a social network may change the 
fate of a commercial product, or just a few super-spreaders 
may determine the virality of a meme in social media.
Despite many recent efforts, the formulation of an accurate
method to optimally identify influential nodes
in complex network topologies remains an unsolved challenge.
Here, we present the exact solution of the problem
for the specific, but highly relevant, case of the 
Susceptible-Infected-Removed (SIR) model for epidemic spreading at
criticality. 
By exploiting the mapping between bond percolation and the static 
properties of SIR, we prove that the recently introduced Non-Backtracking
centrality is the optimal criterion for the identification of
influential spreaders in locally tree-like networks at criticality.
By means of simulations on synthetic networks and on a very extensive set 
of real-world networks,
we show that the Non-Backtracking centrality is a highly reliable 
metric to identify top influential spreaders also 
in generic graphs not embedded in space,  
and for noncritical spreading.
\end{abstract}


\maketitle

\section{Introduction}

Social, technological and biological systems are often characterized
by underlying interaction topologies with complex
features~\cite{albert2002statistical, newman2010networks}.
In a complex network, the roles played by individual nodes 
are highly heterogeneous. Understanding the impact of individual vertices
on the global functionality of the system
is one of the most fundamental, yet not fully solved, 
problems of network science. Centrality measures have 
indeed the purpose
of quantitatively gauging the importance of individual vertices~\cite{wasserman1994social}.
Among the most natural and used ones are degree,
betweenness centrality~\cite{freeman77}, k-shell (or k-core) index~\cite{Seidman1983269},
and eigenvector centrality~\cite{Bonacich72}.

Spreading is at the root of a vast class of phenomena occurring on
network substrates: the propagation of contagious diseases~\cite{PastorSatorras15}, the
diffusion of information or memes~\cite{ratkiewicz2011truthy}, the adoption 
of innovations~\cite{bakshy2009social}, etc.
A large interest has been recently devoted to the
identification of influential spreaders (often called
super-spreaders), i.e., 
nodes that, if chosen as initiators, maximize
the extent of a spreading process.
The goal is to identify which of the many centrality metrics, that
can be computed using only topological information, 
is most strongly correlated with the ability of a node 
to originate massive spreading events. 
%Those metrics should 
%be also scalable, so that they can be computed in a reasonable 
%time even in
%extremely large networks.

Probably, a fully universal method, able to perfectly single out
the most influential nodes for arbitrary spreading dynamics on 
arbitrary networks, does not exist. It is in fact reasonable to expect
that the predictive power of the different centralities strongly 
depends not only on the topology of the underlying network but also
on the details of the spreading process. 
Numerical evidence in this sense can be found 
in~\cite{Borge-Holthoefer12,Ferraz14}.
A much more reasonable goal is instead to identify a metric
able to optimally solve the problem for specific types of dynamics.
Here, we take this path and concentrate our attention on the 
Susceptible-Infected-Removed (SIR) 
model for epidemics. SIR is a paradigmatic model for spreading,
and the vast majority of the investigations about the identification
of influential spreaders in complex networks have dealt with it. 
In random networks, classical results on the SIR model relate the 
epidemic threshold to moments of the degree
distribution~\cite{PastorSatorras15}.
Hence, a naive hypothesis is to assume that the spreading 
ability is strongly correlated to the degree of 
the initiator. 
This view has been challenged by Kitsak {\it et al.}, who
proposed the k-core (also called k-shell) index (which singles out nodes
belonging to dense, mutually interconnected, subgraphs) 
as a proper indicator of the spreading ability~\cite{Kitsak10}. 
This seminal paper has been followed by an avalanche of other 
studies aimed at investigating 
the issue for the same or different dynamical processes, synthetic or 
real-world networks, using a wide range of centralities proposed as
predictors of the spreading ability of the different 
vertices~\cite{Bauer12,klemm2012measure,Chen12,DaSilva12,Chen13,Liu13,Ren14,Ferraz14,Liu15a,Liu15b,Liu16}.
Many empirical investigations have casted doubts
on the ability of the k-shell index to identify
influential spreaders in various
topologies~\cite{DaSilva12,Liu15a}.
However in a very recent work, Ferraz de Arruda {\it et al.}
have reaffirmed the superiority of the k-shell index and degree
centrality as predictors for top influencial spreaders 
in nonspatial networks~\cite{Ferraz14}. The authors of this paper
proposed also an additional centrality metric, the so-called
generalized random walk accessibility, to overcome limitations of the
k-shell index in spatially embedded networks.
The picture emerging from all these efforts is not satisfactory:
All heuristics proposed are motivated based on physical intuition
but involve uncontrolled approximations; No exact result is available
even for synthetic idealized but nontrivial topologies. Methods are
generally validated numerically on a very limited number of networks,
with no complete control of their topological properties.
In this paper we fill this gap, presenting a physically grounded method
which solves exactly the problem in a nontrivial case, and performs
very well in a very broad spectrum of situations.

Our work is based on the connection existing 
between bond percolation 
and the static properties of the SIR model for 
epidemics~\cite{grassberger1983critical, Newman02,PastorSatorras15}.
Very recent results have pointed out
the crucial role played by the spectral properties of the Non-Backtracking (NB)
matrix in determining the properties of the bond percolation process
in complex networks~\cite{Karrer14,Hamilton14,Radicchi15}.
Combining these two well established facts,
we therefore propose 
the NB centrality~\cite{Martin14} as the quantity of
choice for the identification 
of influential spreaders in disordered topologies.
In particular we show that, on locally tree-like networks,
the NB centrality provides the exact solution to the problem of finding
the best single influential spreader, if critical spreading is considered.
We complement this result with a thorough empirical investigation
of the problem on a very large set of real-world topologies, of social, 
technological and biological origin, exhibiting a large variety of 
size, sparsity, heterogeneity and other topological features.
We compare the performance, as predictors of the spreading power 
of single nodes, of the most important centralities proposed so far.
We show that NB centrality turns out to be, in the majority of cases,
the best quantity able to single out the most influential
initiators of spreading processes in networks.

\section{The problem of influential spreaders}

\subsection{The spreading dynamics}

To model the spreading dynamics, we consider the SIR model, 
the simplest and most studied dynamics for epidemics in the presence of 
acquired immunity~\cite{PastorSatorras15}.
Each vertex of a network can be either in state S (susceptible), I 
(infected) or R (interpreted either as recovered or 
removed).
We consider the continuous time version of the dynamics.
At each instant of time,
two elementary events may occur: 
(i) \ce{I ->[\nu] R}, 
meaning that,
at rate $\nu$, a spontaneous recovery/removal
 event may turn a node in state I
into state R;
(ii) \ce{I + S ->[\beta] 2I}, indicating the spreading of the
infection, at rate $\beta$, among pairs of connected 
nodes in states I and S.
Starting from an initial configuration
where all vertices are in the state S and 
only node $i$ is in state I, a 
connected set of 
contiguous vertices may be infected, but after some time all 
infected nodes eventually switch to the R state and the outbreak ends.
The total number $Q_i$ of nodes whose final state is R represents
the extent of the spreading event originated by the single seed $i$.
The problem of interest here is the identification of influential spreaders,
i.e., finding, based only on the topology of the network, what
node of the network must be selected as an initiator of the epidemics
in order to
maximize the average outbreak size.
The asymptotic behavior of the SIR model depends 
on the ratio $\lambda=\beta/\nu$.
If initiators are randomly chosen, then one can define
a critical threshold $\lambda_c$.
For $\lambda$ smaller than the epidemic 
threshold $\lambda_c$, spreading events are of finite (subextensive) size. 
For $\lambda>\lambda_c$ instead the
infection involves a finite fraction of the whole system. 
It is therefore reasonable to expect that also the identity and
role of influential spreaders depends on the value of $\lambda$. 
See the Appendix for info about how $\lambda_c$ is determined numerically.

\subsection{Numerical simulations}
For a given network, we rank 
the nodes on the basis of their spreading power.
We numerically simulate the SIR dynamical process 
with a single initial seed $i$
in state I and  all other nodes in state S. After the dynamics
has ended, we record the number $Q_i$ of nodes in state R.
We then repeat the procedure $10^4$ times,
and quantify the spreading power of node $i$ as 
$\langle Q_i \rangle$, that is the average size of the outbreak
generated from the initial seed $i$.
The measure  $\langle Q_i \rangle$ and its associated 
ranking are the benchmarks against which we compare
the centralities proposed to identify influential spreaders.
We consider four standard centrality metrics: degree, 
k-core, eigenvector centrality, 
and the generalized Random Walk Accessibility (RWA). 
Eigenvector centrality has been indicated
as an effective predictor within mean field analyses, i.e. 
neglecting dynamical correlations between states of 
adjacent vertices~\cite{klemm2012measure}.
RWA has been recently
identified by Ferraz de Arruda {\it et al.}  as the best
predictor for influential spreaders
in spatial networks~\cite{Ferraz14}.
Quantitative comparisons of performance 
among centrality metrics are based on two 
complementary  measures:  the imprecision 
function $\epsilon$~\cite{Kitsak10},  and the Jaccard distance $d_J$. 
Both measures take as input two sets of nodes.
The first is the list of the first $\rho N$ actual top spreaders, 
with $0 < \rho \leq 1$ and $N$ size of the network, as
identified from the results of numerical simulations of the SIR model,
hence ranked on the basis of the score $\langle Q_i \rangle$.
The second set is the list of $\rho N$ top nodes when nodes are
ranked according to the centrality score $x_i$. 
Both $\epsilon$ and $d_J$ return a value ranging between 0, 
for perfect matching
(i.e. the centrality $x$ perfectly predicts the spreading influence
of the fraction $\rho$ of top spreaders) and 1, for completely
failed prediction.
The complementarity between the two measures of performance is apparent 
from their definitions (see Appendix). 
The Jaccard distance measures the difference
among the ``identity'' of the nodes included in the sets of true and 
predicted $\rho N$ top influencers.  
The imprecision function is completely insensitive to
the identity of the nodes, and is determined instead only by 
their spreading power.

\section{Results}

\subsection{Exact solution on locally tree-like networks: 
Non-Backtracking centrality}
The Hashimoto or Non-Backtracking (NB) matrix
is a special representation of the structure of a
network~\cite{Hashimoto89}.
In an arbitrary undirected and unweighted network with $E$ 
edges, the NB matrix is a $2E \times 2E$ array defined
as follows. Every edge $i \leftrightarrow j$ is split
in two directed edges $i \to j$ and $j\to i$.  
The generic entry of the NB matrix
is $M_{i\to j,l\to m}  = \delta_{j,l} (1 - \delta_{i,m})$,
 where $\delta_{i,j}$ is the Kronecker symbol. 
$M_{i \to j, l \to m}$ is different from zero and 
equals one only if the edges $i\to j$ and $l\to m$
define a non-backtracking path of length two.
$M$ is an asymmetric matrix with real and positive principal eigenvalue.
The components of the principal eigenvector, namely 
$v_{i\to j}$, can be used to define the NB centrality of 
vertex $i$ as~\cite{Martin14}
\be
n_i = \sum_j A_{ij} v_{i\to j}.
\label{NBcentrality}
\ee
This centrality is similar to the common eigenvector centrality, but 
it disregards the contribution of vertex $i$ to the centrality of its
neighbors, thus avoiding the self-reinforcement effect responsible
in some cases for the localization of the eigenvector 
centrality~\cite{Martin14,PastorSatorras16}.
We remark that the computation of the principal
eigenpair of the matrix $M$ can be performed using a
simple power-iteration method. 
This allows to estimate the NB centrality of all nodes
in a time that scales as $\mathcal{O}(E)$.
The Ihara-Bass determinant formula may be further used to
reduce memory storage in the computation of the
NB centrality~\cite{bass1992ihara}.

The NB matrix has been shown to play a crucial
role in the problem of graph-clustering~\cite{Krzakala13} and, 
more recently, in percolation~\cite{Karrer14, PhysRevLett.113.208701}. 
In particular,
the percolation threshold in locally tree-like networks is exactly
given by the inverse of the largest eigenvalue of the 
NB matrix for both bond and site 
percolation~\cite{Karrer14, PhysRevLett.113.208701, Radicchi15}. 
As a consequence, the probability that node $i$ is part of the
percolating cluster immediately above the threshold is given by
the expression of Eq.~(\ref{NBcentrality}).

\begin{figure}[!htb]
\includegraphics[width=0.45\textwidth]{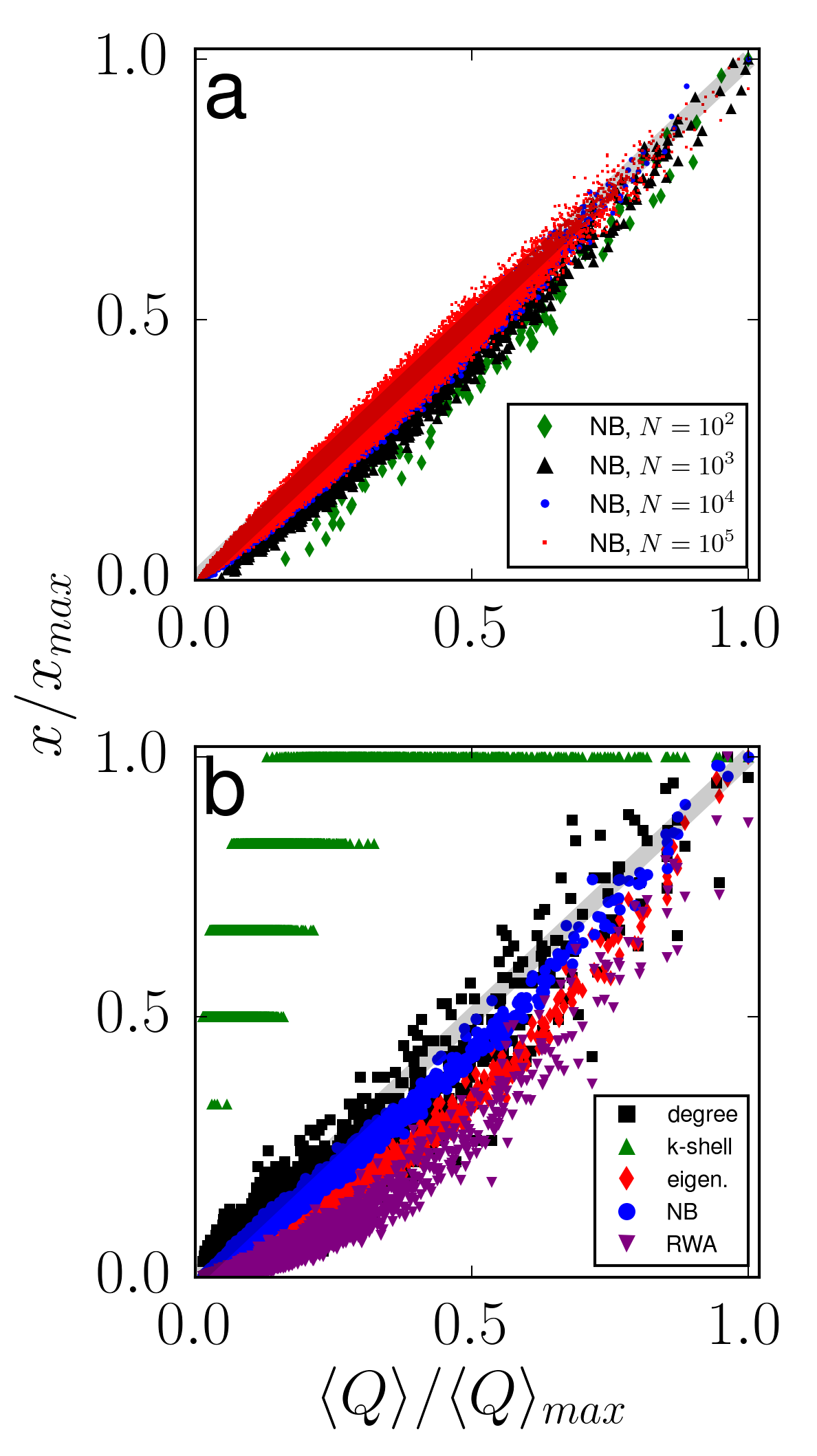}
\caption{
Impact of individual nodes at criticality.
The scatter plot shows the relation between the predicted impact
$x_i$ and the actual impact in simulations $\langle Q_i \rangle$.
Both measures are divided by their maximal values to 
obtain numbers in the interval $[0, 1]$.
Predicted impact is determined on the basis of the centrality
score associated to every node. 
Panel (a) refers to an Erd\H{o}s-R\'enyi (ER) graph with average 
degree $\langle k \rangle  = 4$ and varying size $N$. Only NB centrality
is considered.
Panel (b) refers to a scale-free graph with $N=10^4$ nodes constructed 
according to the uncorrelated configuration model~\cite{Catanzaro05}. 
The degree sequence is composed of integer numbers
selected randomly from a probability distribution 
$P(k) \sim k^{-\gamma}$ defined over the interval $[3, \sqrt{N}]$. 
We consider here $\gamma = 2.5$.
}
\label{scatter}
\end{figure}

The mapping between the static properties of the SIR model and bond
percolation~\cite{Grassberger83,Newman02,PastorSatorras15}
reveals that SIR epidemic outbreaks coincide with the clusters 
of the associated bond percolation process, where the bond occupation
probability $p$ for percolation and the effective spreading rate for SIR
are related by $p=\lambda/(1+\lambda)$. 
This connection has a very important consequence: 
the relative size of an epidemic outbreak started from a specific node $i$
is proportional to the probability that $i$ belongs to the percolating cluster. 
At the critical point, this probability coincides with the NB centrality,
thus
\be
\langle Q_i \rangle \propto n_i.
\label{connection}
\ee
As a consequence the top spreaders are the vertices with the 
highest NB centrality.
Note that this is an exact result, provided the network structure 
is locally tree-like~\footnote{The mapping is strictly valid for a SIR
model where the recovery time is fixed. For the SIR version considered 
in simulations here, the recovery time is nondegenerate and this
implies that the mapping is not strictly
exact~\cite{Meyers2006,Miller2007,Kenah2007}.}.
In Fig.~\ref{scatter} we test numerically the validity of this connection
in synthetic networks.
Panel (a) confirms that Eq.~(\ref{connection}) is generally well obeyed
and tends to be more and more precise as the system size grows, thus making
the network more and more locally tree-like.
Panel (b) shows instead that the values of the other node centralities
have a lower degree of proportionality with the average outbreak size
initiated by them.

The exact equivalence between SIR outbreak size and NB centrality
holds only at criticality, i.e. for $\lambda_c=p_c/(1-p_c)$.
As we depart from $\lambda_c$ the equivalence becomes less accurate.
The probability of belonging to the percolating cluster is no more 
equal to $n_i$. Moreover, below $\lambda_c$ the largest 
cluster does not dominate the cluster size distribution of percolation.
We stress however that criticality is the regime where the
identification of influential spreaders really matters:
The further we move away from the critical point, 
the less interesting and nontrivial the problem becomes.
For large values of $\lambda$ in the supercritical regime, 
any seed will lead to large outbreaks involving a very large
portion of the entire network.
In the deeply subcritical regime instead, at 
very low values of $\lambda > 0$, all spreading events involve
a very small neighborhood of the initial seed.
Only around criticality the choice of the initiator may have
substantial impact on the spreading event, i.e., 
whether the spreading phenomenon remains confined to a few nodes
or it reaches an extensive fraction of the network.

\subsection{Top spreaders in synthetic networks}

We now test the implications of the results in the previous section
for spreading on locally tree-like synthetic networks. 
\begin{figure}
\includegraphics[width=0.45\textwidth]{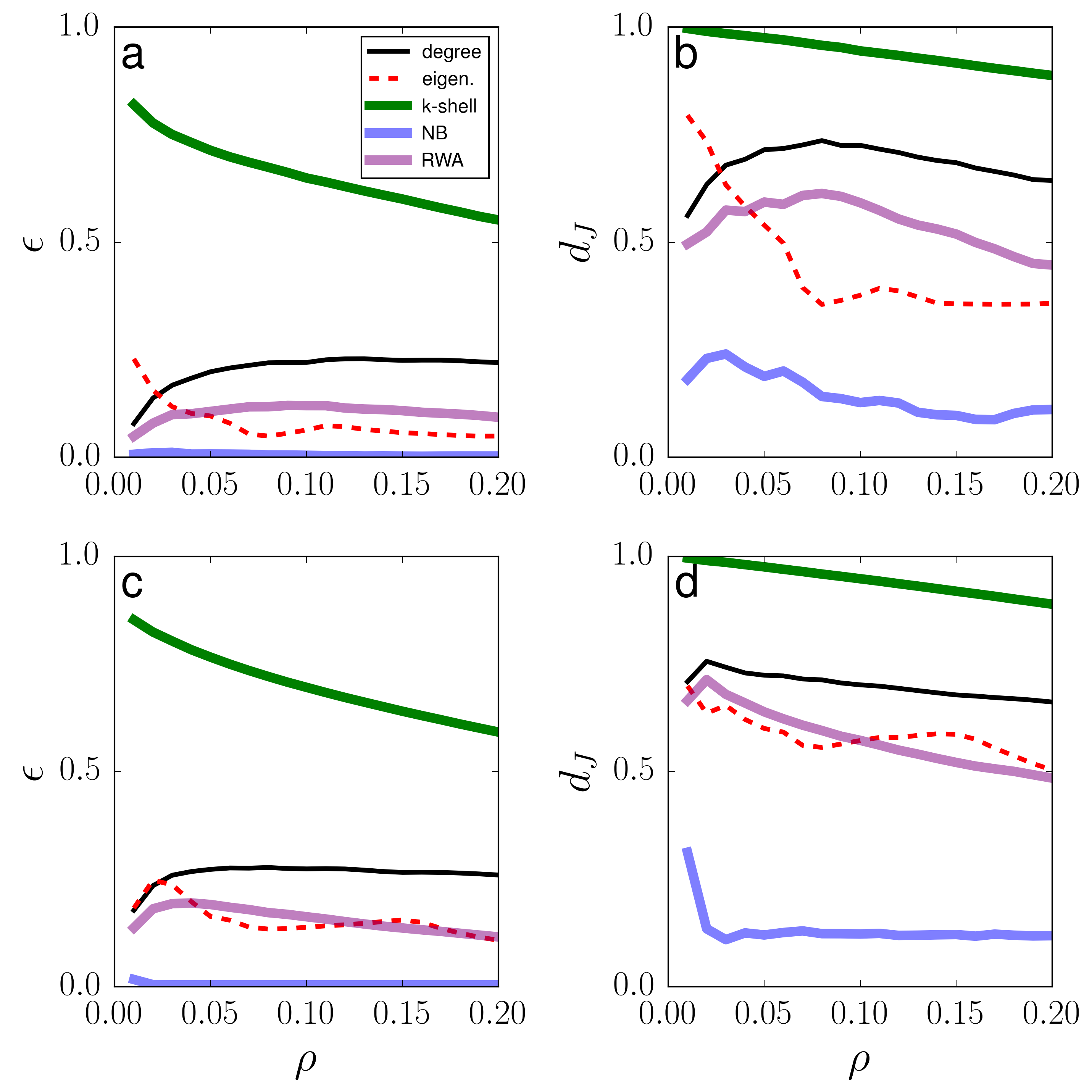}
\caption{\label{syntheticcritical2}
Identification of influential spreaders in Scale-Free (SF)
graphs at criticality. The imprecision function 
$\epsilon$ (panels a and c) and the Jaccard distance
$d_J$ (panels b and d) 
are plotted against the fraction of top nodes $\rho$.
Relative performance of the various centrality measures 
as a function of $\rho$ can be deduced from the direct comparison among
the curves: the lower is the value of the dissimilarity metrics, the
better the centrality measure predicts true top spreaders.
Results are obtained on 
the largest connected component of SF graphs, constructed 
according to the uncorrelated configuration model~\cite{Catanzaro05}.
The pre-imposed degree sequence is composed of random integer numbers
selected randomly from a probability distribution $P(k) \sim k^{-\gamma}$
defined over the interval $[3, \sqrt{N}]$. We consider the case
$\gamma = 3.5$, and two distinct network sizes:
$N=10^4$ (panels a and b)
and $N=10^5$ (panels c and d). 
Numerical simulations
of the SIR model are performed at the critical 
values of $\lambda$ ($\lambda_c = 0.216$ in panels a and b, and 
$\lambda_c = 0.212$ in panels c and d).
}
\end{figure}
We consider a network with degree distribution decaying as
$P(k) \sim k^{-\gamma}$, with $\gamma=3.5$, and compare the
performance of the various centralities as predictors of the
top spreaders in the network.
In Fig.~\ref{syntheticcritical2} we plot the two dissimilarity measures
$\epsilon$ and $d_J$, for the various centralities, as a function of
the fraction $\rho$ of top-ranking nodes. The imprecision function
$\epsilon$ provides a very clear picture: the outbreaks started in
nodes with highest NB centrality are of the same size as those
initiated by the best influential spreaders in numerical simulations.
The degree, the eigenvector centrality, the generalized random walk
accessibility, and, most markedly, the k-shell
index perform much worse.
The plot for $d_J$ gives a similar message, with the difference that
the measure does not vanish for the NB centrality. 
This last observation can be understood by considering
that the NB centralities of distinct nodes do not differ 
much (see Fig.~\ref{scatter}).
Therefore, it is likely that small uncertainties in the
values of $\langle Q_i \rangle$ calculated numerically
may considerably alter the ranking of the nodes, leading to
a nonvanishing Jaccard dissimilarity. 
For the very same reason,  the numerical uncertainties have no 
appreciable effect on the imprecision function $\epsilon$, which 
is very close to 0.

If the same analysis is repeated for other values of the exponent $\gamma$
or for Erd\H{o}s-R\'enyi networks, a very similar phenomenology is found
(see SM1 and SM2): NB centrality outperforms 
eigenvector centrality and generalized random walk accessibility 
in the identification of influential spreaders.
Degree and k-core centrality still deliver poor performances.

We conclude that NB centrality is the optimal choice for 
the selection of influential spreaders on locally tree-like 
networks at criticality. 
The same considerations extend to the subcritical and supercritical 
regimes, provided that $\lambda$ is not too far away from the critical
point (Fig.~SM3).

\subsection{Top spreaders in real-world networks}

\begin{figure}
\includegraphics[width=0.45\textwidth]{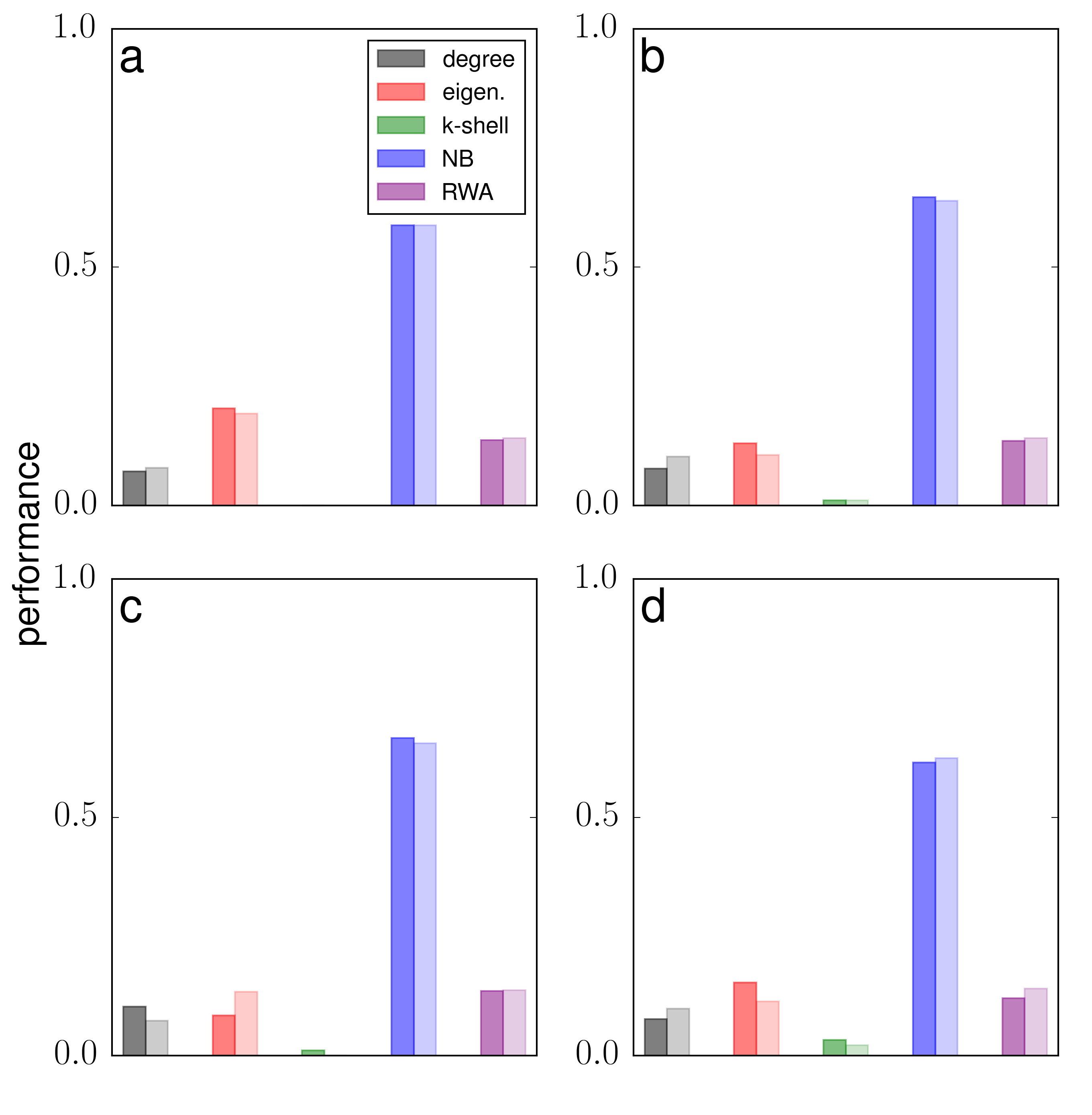}
\caption{\label{compactcritical}
Comparison of predictive power of the different centralities
in nonspatially embedded real-world graphs at criticality. 
The bars indicate the 
fraction of real networks where each centrality
provides the best prediction for the top $\rho N$ influential 
spreaders. 
For every network in our sample, we determine 
the best centrality measure as the metric generating 
the minimal value of
the imprecision function (dark-shaded bars)
or the Jaccard distance (light-shaded bars) 
at predetermined values of $\rho$.
We consider $\rho = 0.05$ (a), $\rho = 0.10$ (b),
$\rho = 0.15$ (c), and $\rho = 0.20$ (d). In the case of 
ties, the score is equally split among the top metrics.   
}
\end{figure}

As substrates for the spreading dynamics, we now consider a very large 
collection of real-world topologies of diverse origin, size and 
topological features. 
Many of these networks have a sizeable clustering coefficient, so that
they cannot be considered, even approximately, as tree-like.
We analyze a total of $95$ networks. Details can be found in the SM.
\begin{figure}
\includegraphics[width=0.45\textwidth]{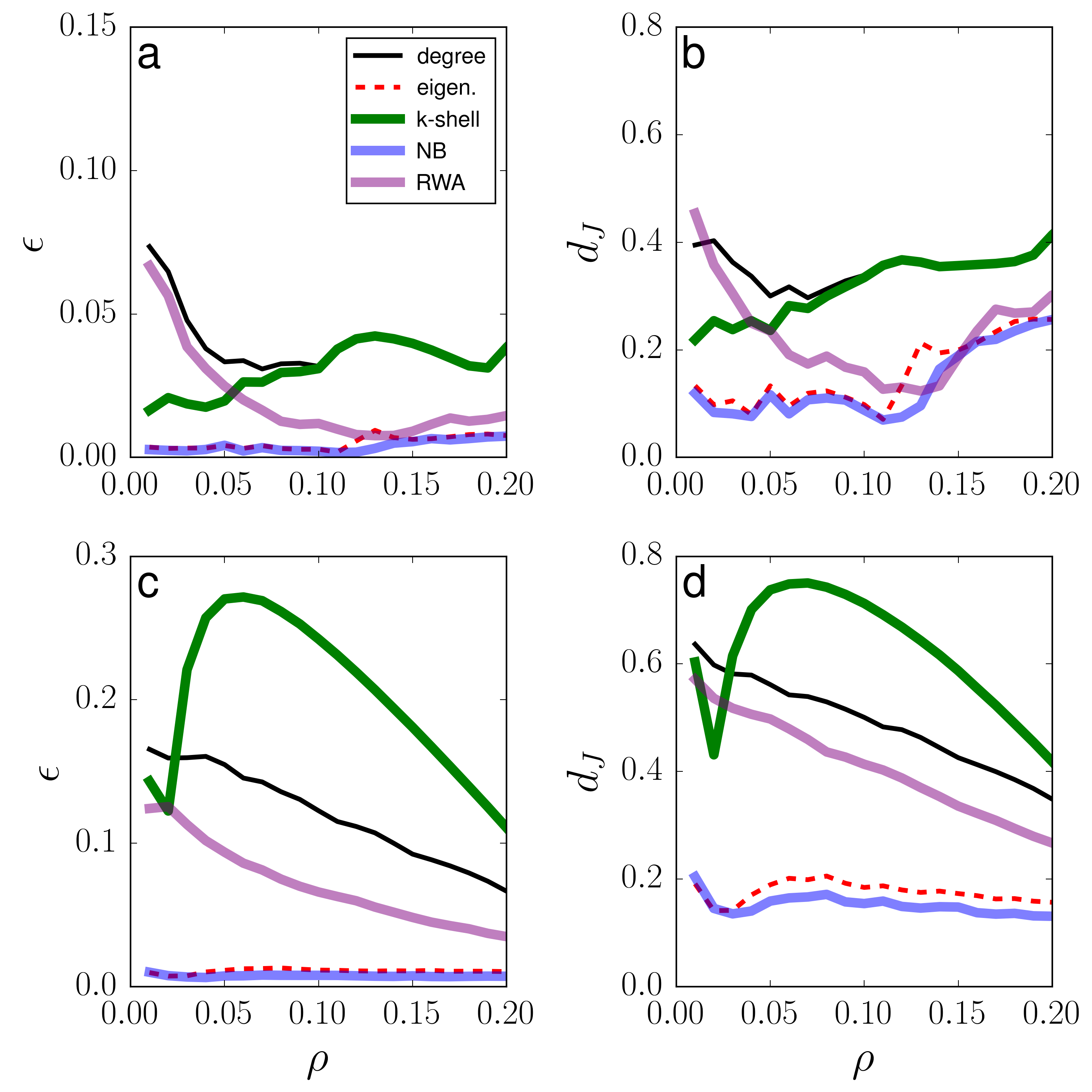}
\caption{\label{critical}
Identification of influential spreaders in real-world
graphs at criticality.
The description of the various panels is similar to those of
Fig.~\ref{syntheticcritical2}.
In panels a and b, we present results for the Email contact
network~\cite{Kitsak10}.
In panels c and d, we show the results for the peer-to-peer network of 
Gnutella as of August 31, 2002~\cite{leskovec2007}.
The clustering coefficients of the networks are 0.1088 and 0.0055, respectively.
SIR simulations are performed at criticality, with $\lambda = \lambda_c$.
The epidemic thresholds of the two networks are 
$\lambda_c = 0.031$ and $\lambda_c = 0.099$, respectively.
}
\end{figure}
In Fig.~\ref{critical}, we present the results for two such
networks: a graph of email contacts~\cite{Kitsak10}, and 
the Gnutella peer-to-peer network~\cite{leskovec2007}.
It turns out very clearly that, for these structures,
k-shell centrality and degree perform very badly; RWA
performs slightly better, but still poorly; eigenvector and NB 
centralities are instead very effective in identifying influential
spreaders.
Among these two, NB centrality provides a slightly 
more effective recipe for the identification of top influential spreaders.
The picture obtained for synthetic networks is then essentially confirmed.
We have repeated the same analysis for a very large set of networks
with nonspatial embedding (Tables~SM1,~SM2, and~SM3).
The set of networks include graphs of different
nature (e.g., biological, technological, social) thus with
large variability in their topological features (e.g., degree
distribution, size, clustering coefficient).
Whereas some variation exists depending on the detailed
topology, overall the message is clear: the NB centrality 
of a node is, in about $60\%$ of the networks analyzed,
the most accurate predictor of the spreading
ability of individual nodes (Fig.~\ref{compactcritical}).
NB centrality outperforms all other centrality metrics
in real nonspatial networks. Only in graphs with spatial embedding
RWA provides better performances (Fig.~SM6).

Previous results are obtained for critical spreading, where the
susceptibility of the system is maximal. We repeat the analysis
for the subcritical regime by setting $\lambda=2/3 \, \lambda_c$, and 
the supercritical phase for $\lambda=3/2\,\lambda_c$ (Figs.~SM4 and~SM5). 
The overall picture is again similar to the one observed for 
critical spreading: the NB centrality is in the majority of cases
the most accurate predictor to identify influential spreaders.

\section{Conclusions}

\begin{figure*}[!htb]
\includegraphics[width=0.95\textwidth]{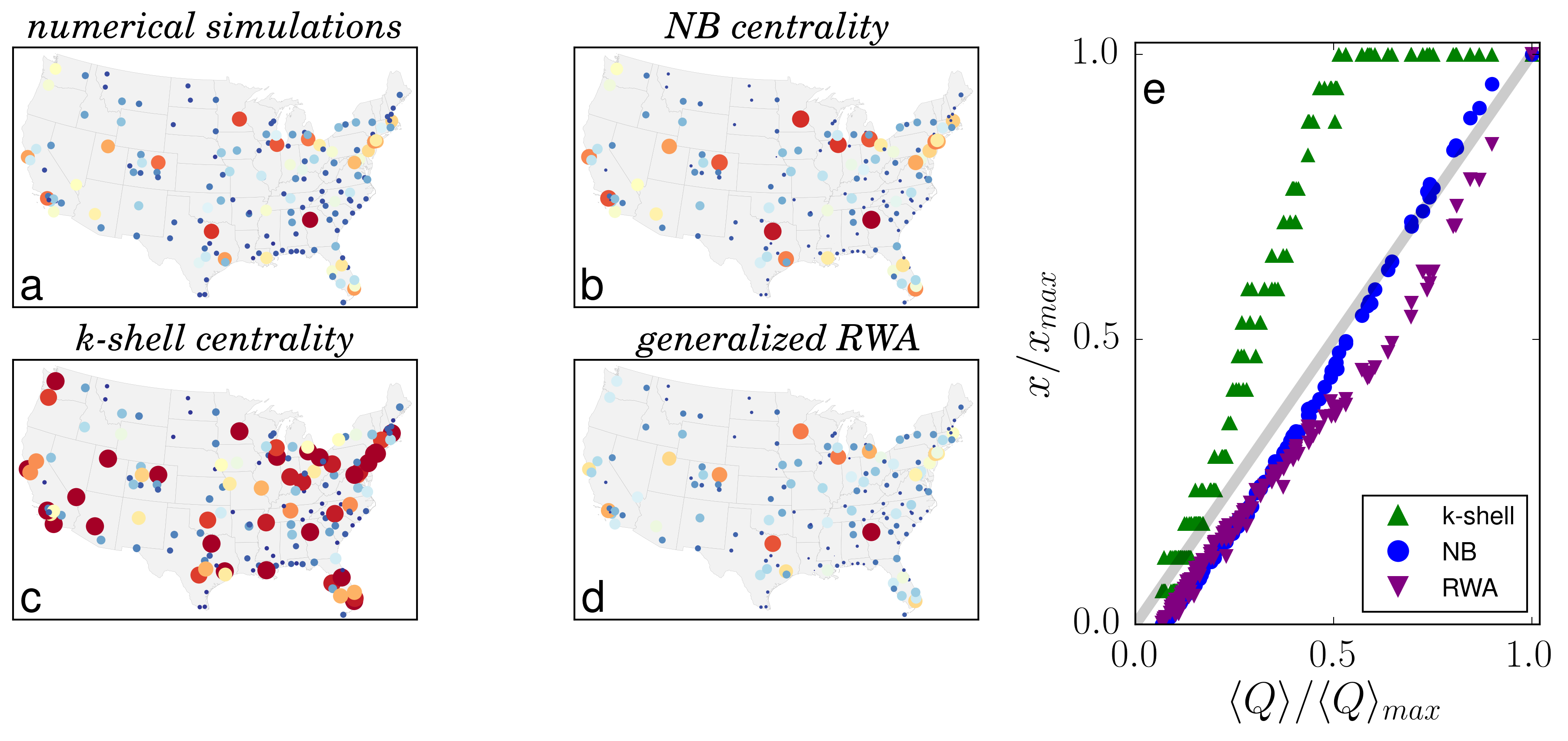}
\caption{\label{airus}
Influential spreaders in the US air transportation network at criticality.
We consider the unweighted and undirected version of the
air transportation network within the US, reconstructed by aggregating
the information of all flights by major 
carriers (American Airlines, Delta and United) in 
January 2014~\cite{flights, radicchi2015percolation}.
Panel a provides a visualization of the
impact of individual airports in the spreading process.
The impact of airport $i$ is measured as $\langle Q_i \rangle /
\langle Q \rangle_{max}$,
with $\langle Q \rangle_{max}$ maximal value of the
spreading power over all airports. In the representation, the
size of the circles is proportional to the impact of the corresponding
airport. Colors of the circles are also proportional to the 
spreading power of the
airport, with a continuous scale ranging from red (maximal impact)
to blue (minimal impact). In panels b and c, we provide the same
representation
as in panel a, but replacing the spreading power with the value
of their centrality. In panel d, we provide a scatter plot
the centrality metrics against the value of the spreading power for
individual airports. The grey line is the benchmark for perfect
performance.
Whereas NB centrality and generalized RWA provide 
a distinction for airports, k-core centrality generates identical scores
for many airports.
}
\end{figure*}

The present analysis provides convincing evidence that the centrality
determined from the Non-Backtracking (NB) matrix of a graph represents
the best predictor for the identification of SIR influential spreaders in
the network.  The choice of this centrality measure is motivated by 
recent theoretical progress in the study of percolation processes in 
arbitrary locally tree-like graphs, and by
the equivalence between the SIR model and the bond percolation
model at criticality.  Even in real networks, where the locally
tree-like ansatz is violated, NB centrality turns out to greatly outperform
other centrality metrics in the task of identifying top
influential spreaders.  We remark also that NB centrality can be computed
in a time that scales almost linearly with the system size, and it is
thus applicable to very large networks.

An additional, interesting, result emerging from our systematic 
analysis of real networks is that k-shell centrality generally provides 
very unsatisfactory performances, not only compared to NB centrality,
but also to degree, eigenvector centrality and generalized random walk
accessibility.  This is at odds with what claimed in the seminal paper
by Kitsak and collaborators~\cite{Kitsak10}, 
and more recently remarked by Ferraz de Arruda {\it et
  al.}~\cite{Ferraz14}, with the analysis
of very small samples of real-world networks.
Given the amount of real-world graphs 
considered in our study, we believe
that our message is conclusive: k-shell index can be easily
outperformed by other simple centrality metrics in the identification of
influential nodes in 
dynamical  processes on  complex networks.
One of the main reasons of the poor ability of the k-core to identify top
spreaders is rooted in the very definition of k-core index,
which necessarily involves a large degeneracy~\cite{DaSilva12,Pei13}.
The metric is not able to make a distinction among top vertices in the 
ranking, since, by definition, $k$ nodes must be tied at the top position 
if $k$ is the maximal value of the k-shell centrality measured in a network. 
This fact is clearly illustrated for artificial
graphs in Fig.~\ref{scatter}, where the k-core index is the same
for very large groups of vertices, whereas
their spreading power is highly heterogeneous.
Similar considerations are valid also for real graphs.
In Fig.~\ref{airus}, we consider SIR on a substrate given by an
air transportation network within the US~\cite{flights}. 
The spreading influence of individual nodes is well reproduced 
by the NB centrality and, more approximately, by the RWA. 
According to the k-shell centrality, several
airports are ranked at the top of the list. The top tier is, however,
composed of airports with fundamentally different values
of their spreading power: for example, 
``Hartsfield-Jackson Atlanta International Airport``, the actual top 
spreader in the network, is tied with ``Wilmington International Airport'', 
despite the latter actually has a spreading power twice smaller than the 
top spreader.

Beyond their applicability to relevant real situations, these results
open new exciting perspectives for other, related, problems.
A first question is the validity of the NB centrality solution for
other types of spreading dynamics, different from the SIR class.
While NB centrality is unlikely to perform well for rumor 
dynamics~\cite{Borge-Holthoefer12,Ferraz14}, the question is open
for more complex modeling frameworks for epidemics, such as 
metapopulations~\cite{Grenfell1997,Colizza2008}.
Secondly, the problem studied here refers to individual spreaders.
Substantially different results may arise in the case of optimal
multiple spreaders, i.e. the identification of the subset of 
network vertices (of a given number of nodes) maximizing the 
extent of a spreading process seeded in all of them at the same time. 
As already noted in Ref.~\cite{Kitsak10}, starting the process 
in the best single spreaders often results in suboptimal propagation, 
because of the overlap among the areas of influence of the best 
individual spreaders.
Finding the best set of multiple spreaders is a different, highly nontrivial,
NP-complete optimization problem~\cite{Kempe03}, for which many 
clever approximation schemes have been 
proposed~\cite{Kempe03,Altarelli13,Morone15}, but a scalable and accurate
general approach is still not available.
The insights provided by the mapping to percolation and the consideration
of the NB centrality may pave the way for further progress also in this context.
Another exciting line of research regards the identification of 
influential spreaders from empirical data on real-world 
spreading phenomena~\cite{Gonzalez-Bailon11,pei2014searching}. 
In this respect, the problem is further complicated by the fact that the
spreading dynamics at the microscopic level are not known a priori and
may contain additional ingredients not included in the simple models
usually considered.

\begin{acknowledgments}
FR acknowledges 
support from the National Science 
Foundation (Grant CMMI-1552487) and 
the US Army Research Office (W911NF-16-1-0104).
\end{acknowledgments}

\appendix

\section{Measures of performance}

\subsection{The imprecision function}

The imprecision function~\cite{Kitsak10} $\epsilon(\rho)$ quantifies
the difference between the average size of the spreading events 
initiated (as single spreaders) by the first $\rho N$ vertices according 
to a given centrality and the analogous size for the actual 
$\rho N$ most efficient spreaders in SIR simulations. 
$N$ is the number of nodes in the network and $0<\rho \leq 1$.
More in detail, let us define as $\Upsilon^{(x)}(\rho)$ the set of the top
$\rho N$ vertices according to the centrality $x$ and $\Upsilon^{(eff)}(\rho)$
the actual top $\rho N$ spreaders, as measured in SIR simulations.
The quantity
\be
Z^{(x)}(\rho) = \frac{1}{N \rho} \sum_{i \in \Upsilon^{(x)}(\rho)}
\langle Q_i \rangle
\label{eq:zeta}
\ee
is the average size of outbreaks originated in the most highly ranked
nodes according to the centrality $x$. If $Z^{(eff)}(\rho)$ is the
same quantity as in Eq.~(\ref{eq:zeta}) but 
computed over the set $\Upsilon^{(eff)}(\rho)$,
the imprecision function is defined as
\be
\epsilon^{(x)}(\rho) = 1 - \frac{Z^{(x)}(\rho)}{Z^{(eff)}(\rho)}
\ee
If the centrality $x$ perfectly identifies the most efficient spreaders,
the imprecision function equals zero. 
High values of $\epsilon^{(x)} (\rho)$ indicate
that the centrality is not a good predictor of the
spreading power of the top $\rho N$ spreaders.
To account for possible ties in the centrality metric $x$, 
we average the imprecision function over at least 
10 realizations of the set $\Upsilon^{(x)}(\rho)$.

\subsection{The Jaccard distance}
The Jaccard distance $d_J(\rho)$ is a measure of the 
dissimilarity between 
two  sets $\Upsilon^{(x)}(\rho)$ and 
$\Upsilon^{(eff)}(\rho)$. This quantity is defined as
\be
d_J^{(x)}(\rho)  = 1 - \frac{|\Upsilon^{(x)}(\rho) \cap \Upsilon^{(eff)}(\rho)|}{|\Upsilon^{(x)}(\rho) \cup \Upsilon^{(eff)}(\rho)|}
\ee
where $|A|$ stands for the number of elements
in the set $A$. Clearly, if the two sets $\Upsilon^{(x)}(\rho)$ and
$\Upsilon^{(eff)}(\rho)$ 
coincide the distance vanishes, while if they have null intersection,
then their distance equals one.

\section{Centrality measures}
We consider the following centrality measures.

\begin{itemize}
\item
Degree centrality. This is the simplest centrality measure 
that can be defined for nodes
in a network. The degree of node $i$ equals the number of neighbors
of vertex $i$ in the network. 

\item 
k-shell centrality. 
A k-core is a subset of nodes composed of vertices that
have at least $k$ neighbors within the set itself.
The k-shell or k-core index of a node equals the largest $k$ value of
k-cores which the node belongs to.

\item
Eigenvector centrality. The score assigned to each node
equals the value of the component of the principal
eigenvector of the adjacency matrix
of the network.

\item The score of node $i$ based on the generalized Random Walk Accessibility (RWA) 
is defined as  $\alpha_i = \exp{ ( - \sum_j W_{i,j} \ln{ W_{i,j} }  ) } $,
where $W_{i,j} = \sum_{q=0}^\infty (P^q)_{i,j} /
q!$, with $P^q$ $q$th power of the random walk transition matrix
of the graph~\cite{Ferraz14}. The exact computation of the 
RWA score for all nodes in the network 
requires the diagonalization of the matrix $P$,
an unfeasible task for medium- and large-size networks. 
Good approximations of RWA scores can be
obtained by means of agent-based simulations of 
the random walk dynamics. Our estimates of RWA 
are based on average values obtained over $10^6$
independent walks of maximal  
length $20$ for every  node in the network.

\end{itemize}

\section{Numerical determination of the epidemic thresholds}

For a given network, we determine the critical value $\lambda_c$
in the following way. 
For a given value of $\lambda$, we start from a configuration
where all nodes are in state S, and one randomly chosen vertex is in
state I. We run the SIR model, and measure the size of the outbreak
$Q$. We repeat the procedure $100,000$ times, every time choosing
at random a node as initial seed of the epidemics, 
and compute the first
and second moment of the size of the outbreak, namely
$\langle Q \rangle$ and $\langle Q^2 \rangle$.
The critical value value $\lambda_c$ is determined from the position
of the peak of the ratio 
$\langle Q^2 \rangle / \langle Q \rangle^2$~\cite{PastorSatorras16b}.
Values of $\lambda_c$ for all networks
analyzed in this paper are reported in Tables~SM1,~SM2, and~SM3.

%\bibliography{Influence}

\clearpage

\onecolumngrid

%%%%%%%%%%%%%%%%%%%%%

\renewcommand{\theequation}{SM\arabic{equation}}
\setcounter{equation}{0}
\renewcommand{\thefigure}{SM\arabic{figure}}
\setcounter{figure}{0}
\renewcommand{\thetable}{SM\arabic{table}}
\setcounter{table}{0}

\begin{figure}
\includegraphics[width=0.45\textwidth]{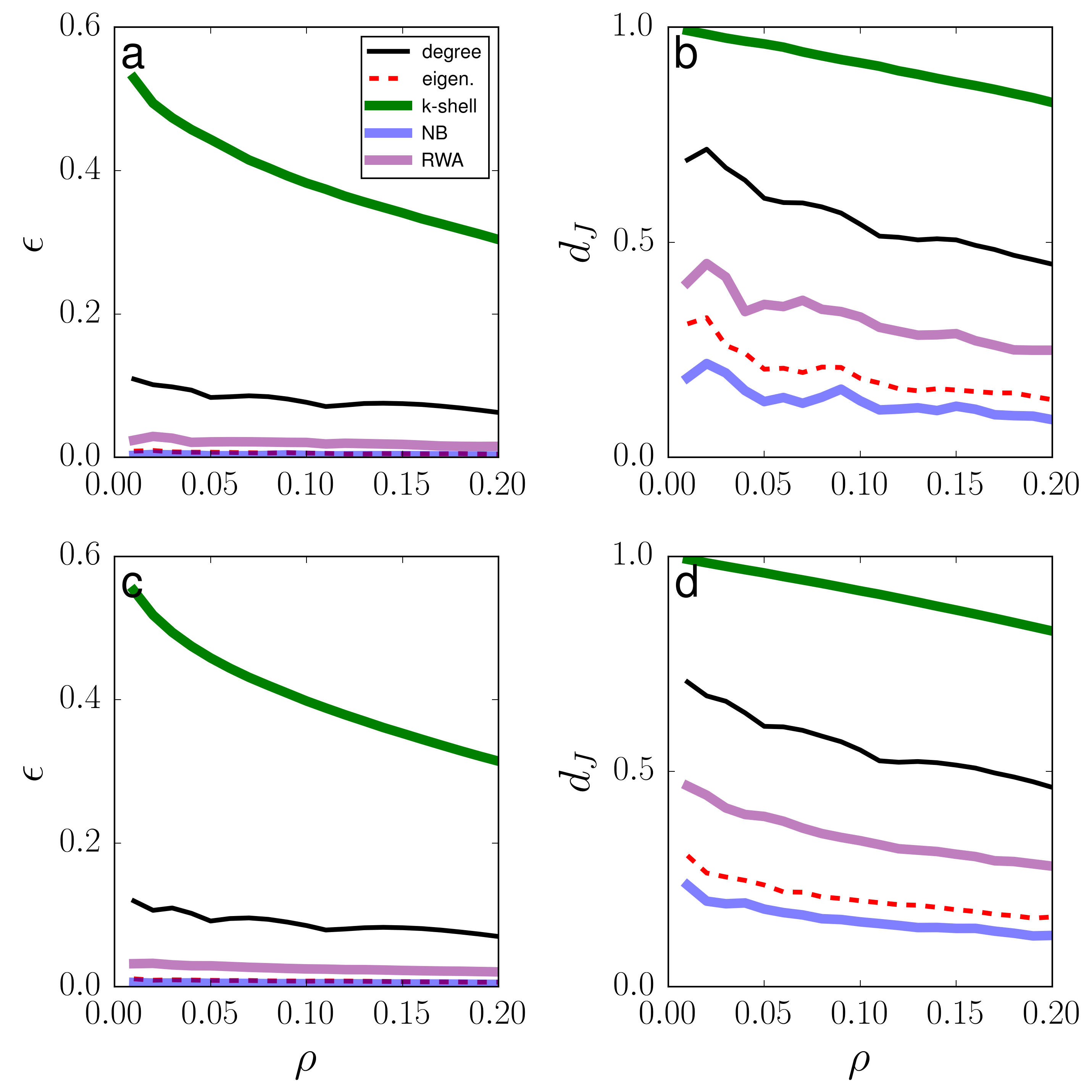}%                                                               
\caption{
Identification of influential spreaders in Erd\H{o}s-R\`enyi (ER)
graphs at criticality. The imprecision function $\epsilon$ (panels
and c) and the Jaccard distance
$d_J$ (panels b and d)
are plotted against the fraction of top nodes $\rho$.
Relative performance of the various centrality measures
as a function of $\rho$ can be deduced from the direct comparison among
the curves: the lower is the value of the dissimilarity metrics, the
better the centrality measure is in the prediction
of true top spreaders.
Results are obtained on the largest connected component of ER graphs
with average degree
$\langle k \rangle = 4$, and size $N=10^4$ (panels a and b)
and $N=10^5$ (panels c and d). Numerical simulations
of the SIR model are performed at the critical
values of $\lambda$ ($\lambda_c = 0.332$ in panels a and b, and
$\lambda_c = 0.331$ in panels c and d).
}
\label{syntheticcritical1}
\end{figure}
\begin{figure}
\includegraphics[width=0.45\textwidth]{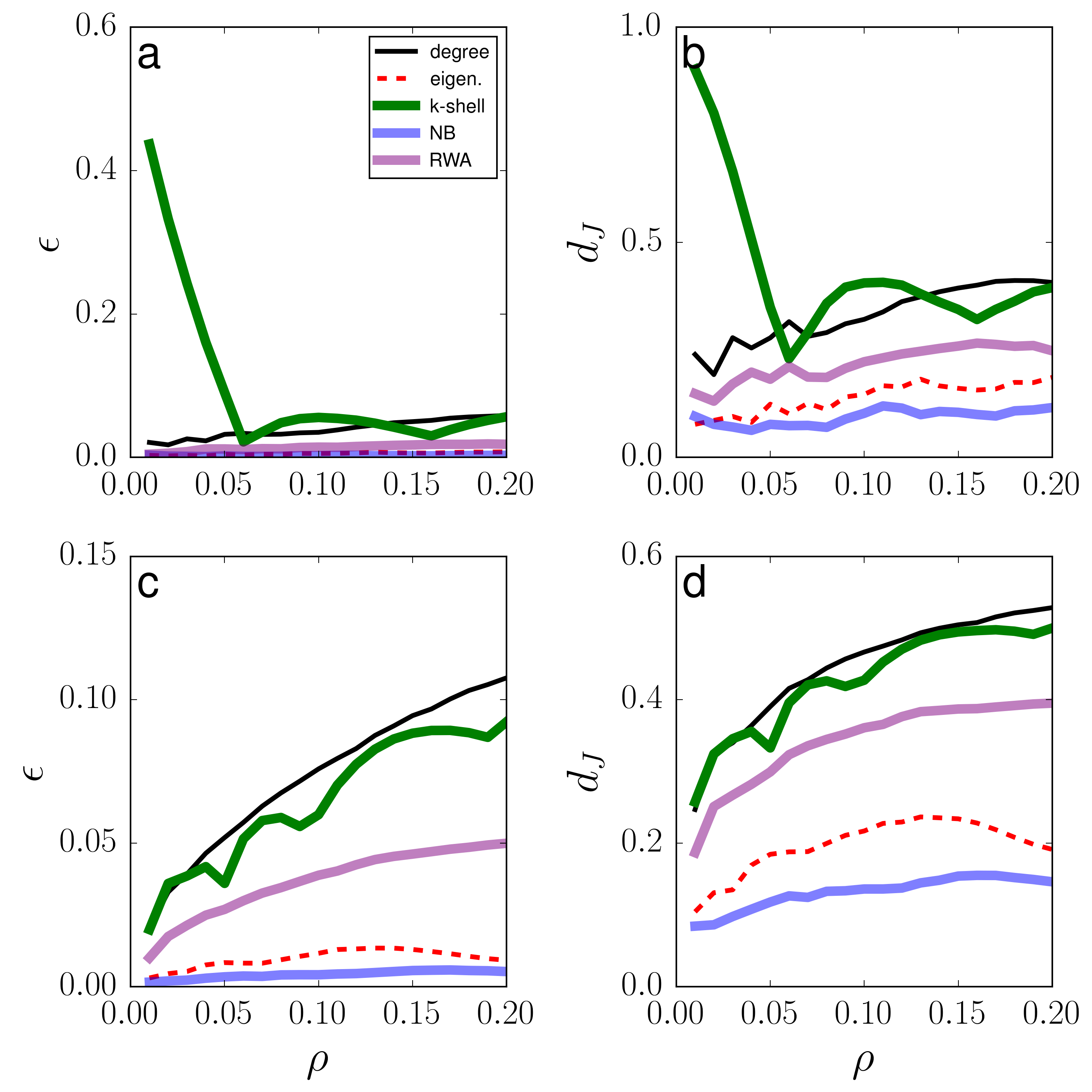}%                                                               
\caption{
Identification of influential spreaders in Scale-Free (SF)
graphs at criticality. The imprecision function 
$\epsilon$ (panels and c) and the Jaccard distance
$d_J$ (panels b and d) 
are plotted against the fraction of top nodes $\rho$.
Relative performance of the various centrality measures 
as a function of $\rho$ can be deduced from the direct comparison among
the curves: the lower is the value of the dissimilarity metrics, the
better the centrality measure is in the prediction 
of true top spreaders.
Results are obtained on 
the largest connected component of SF graphs, constructed 
according to the uncorrelated configuration model. The pre-imposed
degree sequence is composed of random integer numbers
selected randomly from a probability distribution $P(k) \sim
k^{-\gamma}$
defined over the interval $[3, \sqrt{N}]$. We consider the case
$\gamma = 2.5$, and two distinct network sizes:
$N=10^4$ (pannels a and b)
and $N=10^5$ (pannels c and d). 
Numerical simulations
of the SIR model are performed at the critical 
values of $\lambda$ ($\lambda_c = 0.074$ in pannels a and b, and 
$\lambda_c = 0.040$ in pannels c and d).
}
\label{syntheticcritical2a}
\end{figure}
\begin{figure}
\includegraphics[width=0.45\textwidth]{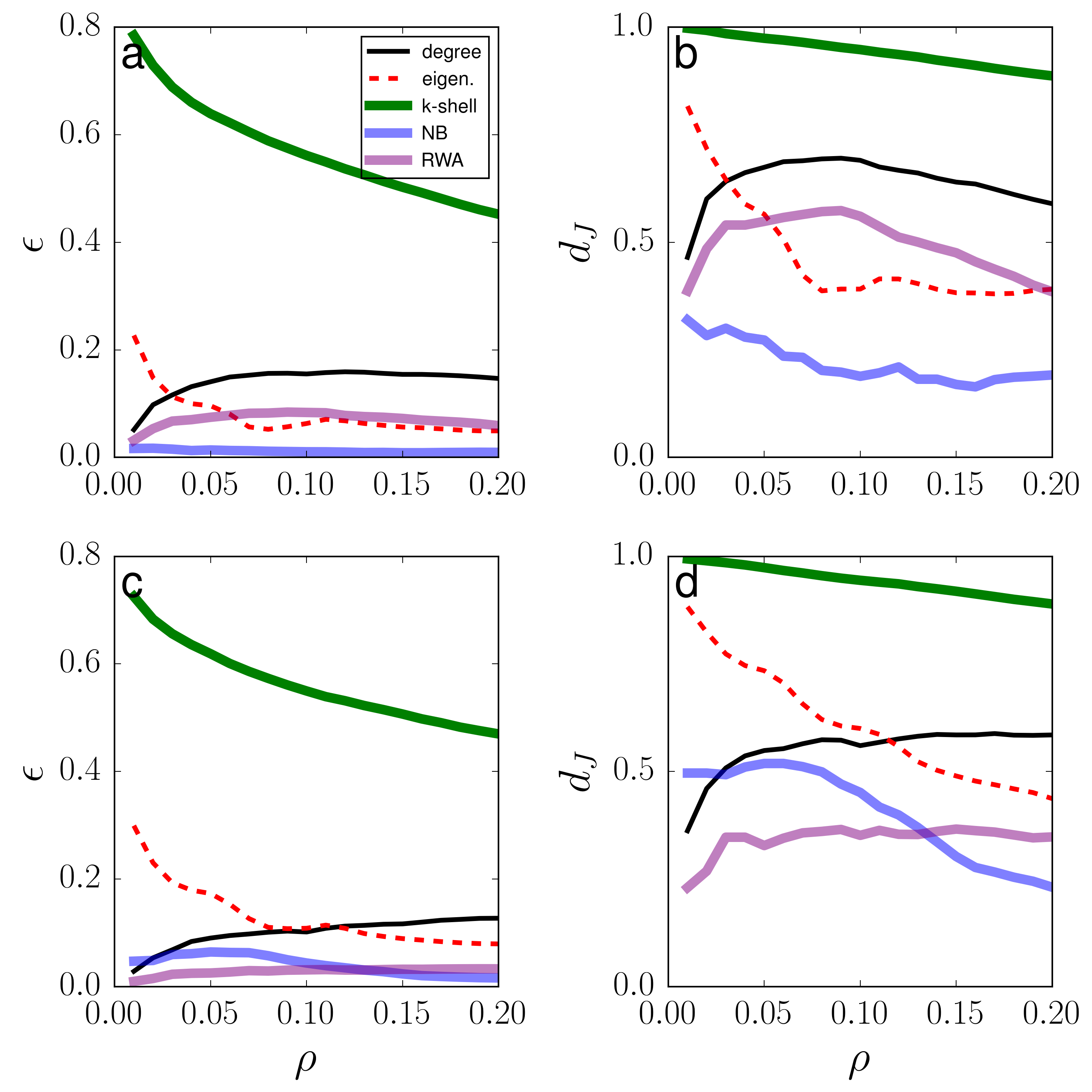}
\caption{\label{syntheticnoncritical}
Identification of influential spreaders in Scale-Free
graphs in the subcritical and supercritical regimes. 
We consider the same network as the one of
Fig.~2a and b of the main text. 
Numerical simulations
of the SIR model are, however, performed off-criticality.
We consider the subcritical regime with $\lambda = 2/3 \, \lambda_c$  
(panels a and b) and the supercritical regime with
$\lambda = 3/2 \, \lambda_c$ (panels c and d).
}
\end{figure}
\begin{figure}
\includegraphics[width=0.45\textwidth]{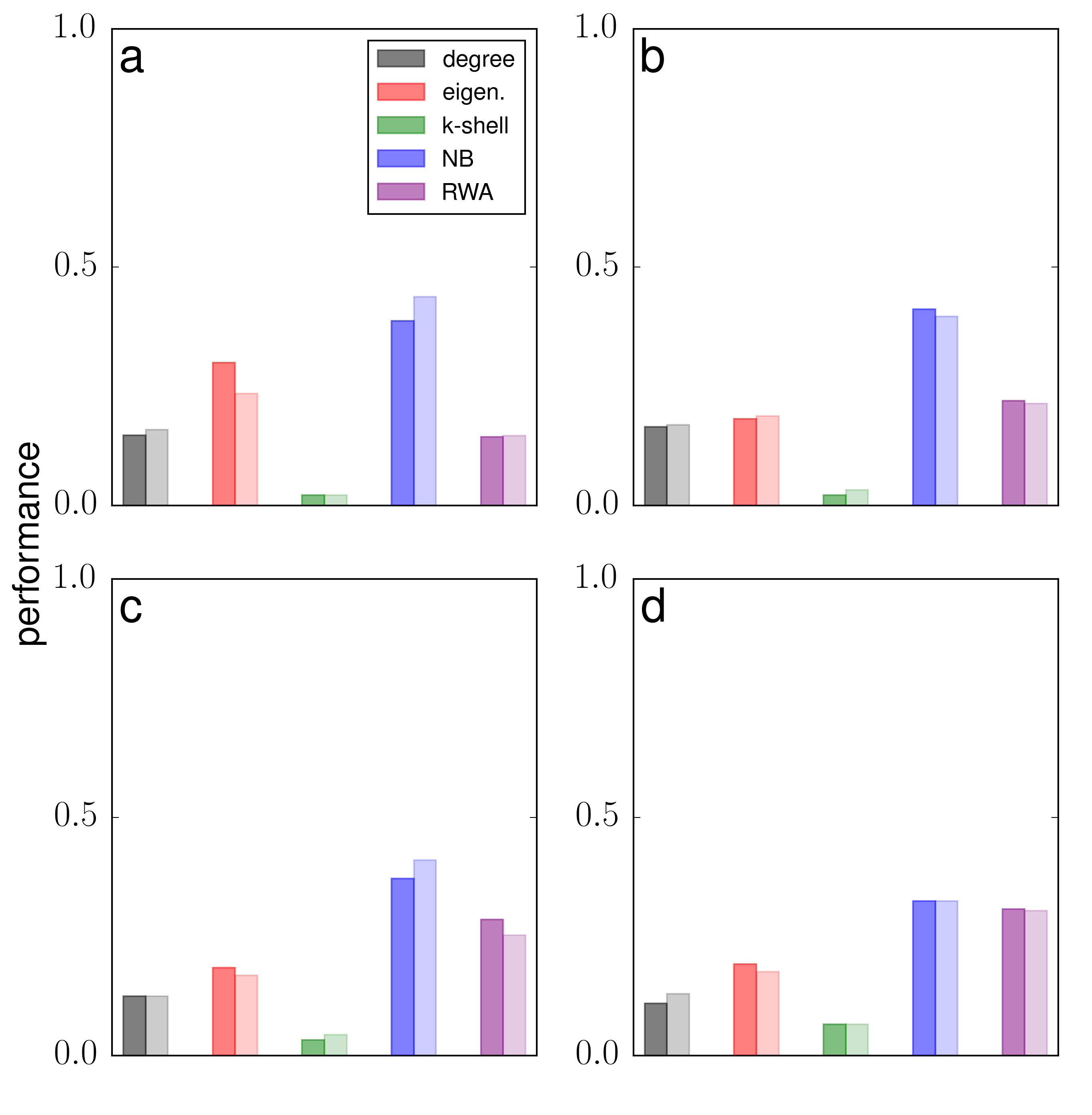}
\caption{\label{compactothers2}
Identification of influential spreaders in real-world networks in the
subcritical regime.
Same as in Fig.~4 of the main text but for subcritical spreading
($\lambda = 2/3 \, \lambda_c$).
}
\end{figure}

\begin{figure}
\includegraphics[width=0.45\textwidth]{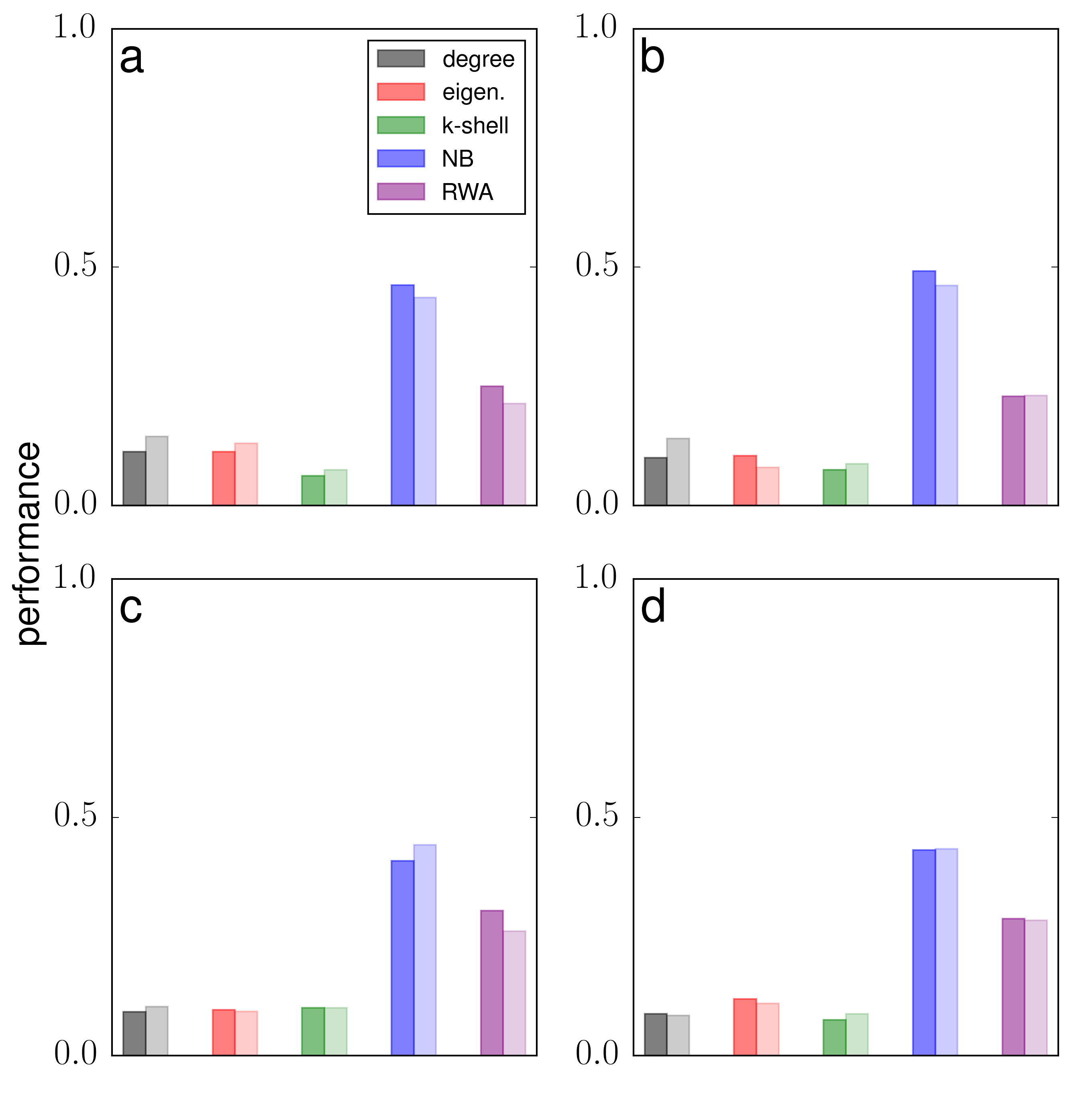}
\caption{\label{compactothers}
Identification of influential spreaders in real-world networks in the
supercritical regime.
Same as in Fig.~4 of the main text but for supercritical spreading
($\lambda = 3/2 \, \lambda_c$).}
\end{figure}

\begin{figure}
\includegraphics[width=0.45\textwidth]{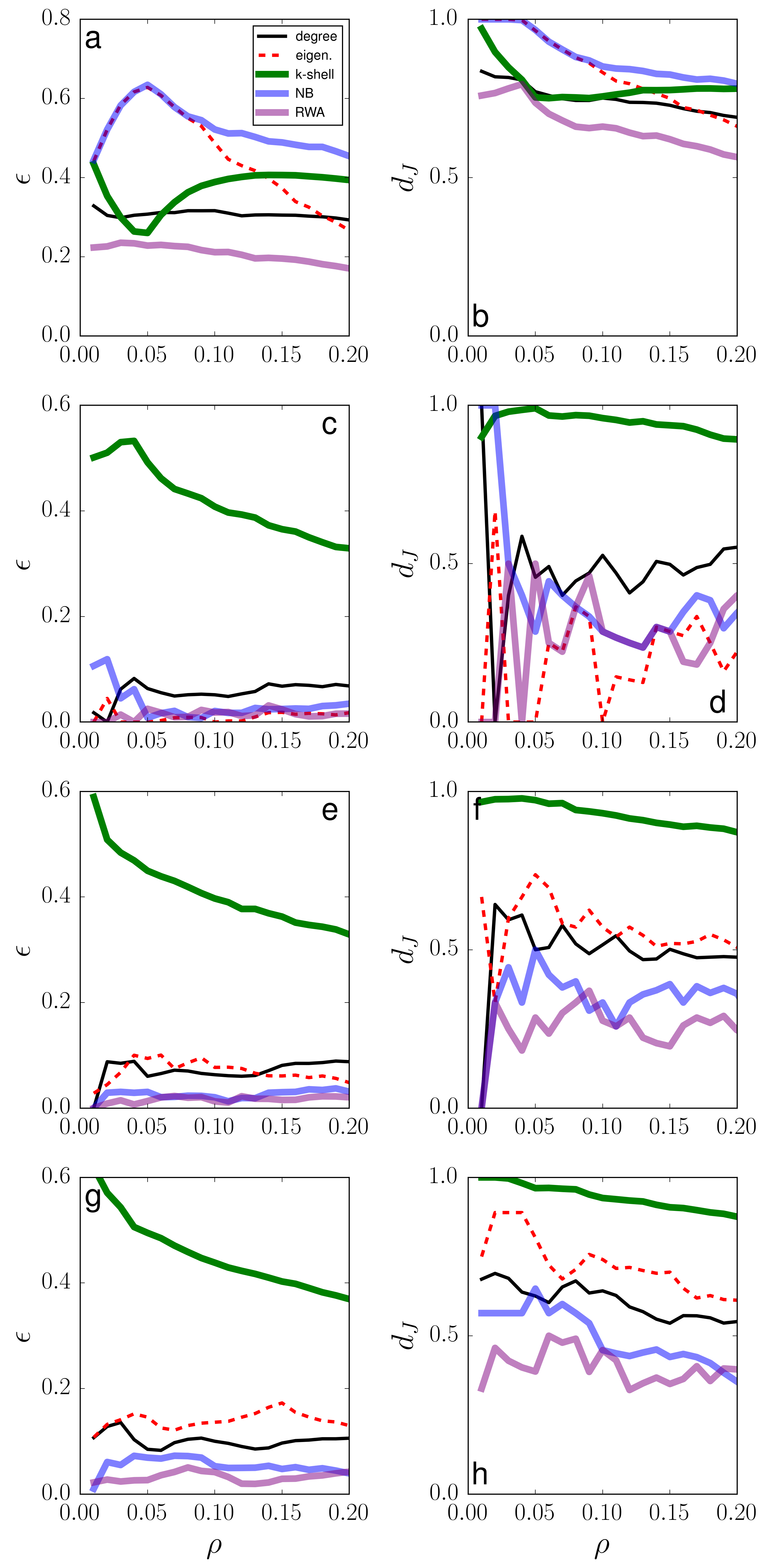}
\caption{\label{spatial}
Identification of influential spreaders in real-world
spatial graphs at criticality.
The description of the various panels is similar to those of
Fig.~3 of the main text.
In panels a and b, we present results for the {\tt US power grid}
network~\cite{watts1998collective}.
In the other panels, we show results for the electronic circuits
networks
{\tt S 208} (c and d), {\tt S 420} (e and f), and {\tt S 838} (g and h)~\cite{milo2004superfamilies}.
}
\end{figure}

\clearpage
\begin{table*}
\begin{center}\begin{tabular}{|r|l|r|r|r|c|c|c|r|r|} \hline
$\#$ & network & $N$ & $E$ & $\lambda_c$ & Sub. & Cri. & Sup.  & Refs. & Url \\ 
 \hline 
$1$&{\tt Social 3} & $32$ & $80$ & $0.241$ & \cmark & \cmark & \cmark & \cite{milo2004superfamilies} &\href{http://wws.weizmann.ac.il/mcb/UriAlon/index.php?q=download/collection-complex-networks}{url} \\ \hline 
$2$&{\tt Karate club} & $34$ & $78$ & $0.219$ & \cmark & \cmark & \cmark & \cite{zachary1977information} &\href{http://www-personal.umich.edu/~mejn/netdata/}{url} \\ \hline 
$3$&{\tt Protein 2} & $53$ & $123$ & $0.293$ & \cmark & \cmark & \cmark & \cite{milo2004superfamilies} &\href{http://wws.weizmann.ac.il/mcb/UriAlon/index.php?q=download/collection-complex-networks}{url} \\ \hline 
$4$&{\tt Dolphins} & $62$ & $159$ & $0.220$ & \cmark & \cmark & \cmark & \cite{lusseau2003bottlenose} &\href{http://www-personal.umich.edu/~mejn/netdata/}{url} \\ \hline 
$5$&{\tt Social 1} & $67$ & $142$ & $0.325$ & \cmark & \cmark & \cmark & \cite{milo2004superfamilies} &\href{http://wws.weizmann.ac.il/mcb/UriAlon/index.php?q=download/collection-complex-networks}{url} \\ \hline 
$6$&{\tt Les Miserables} & $77$ & $254$ & $0.128$ & \cmark & \cmark & \cmark & \cite{knuth1993stanford} &\href{http://www-personal.umich.edu/~mejn/netdata/}{url} \\ \hline 
$7$&{\tt Protein 1} & $95$ & $213$ & $0.402$ & \cmark & \cmark & \cmark & \cite{milo2004superfamilies} &\href{http://wws.weizmann.ac.il/mcb/UriAlon/index.php?q=download/collection-complex-networks}{url} \\ \hline 
$8$&{\tt E. Coli, transcription} & $97$ & $212$ & $0.362$ & \cmark & \cmark & \cmark & \cite{mangan2003structure} &\href{http://wws.weizmann.ac.il/mcb/UriAlon/index.php?q=download/collection-complex-networks}{url} \\ \hline 
$9$&{\tt Political books} & $105$ & $441$ & $0.109$ & \cmark & \cmark & \cmark & \cite{adamic2005political} &\href{http://www-personal.umich.edu/~mejn/netdata/}{url} \\ \hline 
$10$&{\tt David Copperfield} & $112$ & $425$ & $0.097$ & \cmark & \cmark & \cmark & \cite{newman2006finding} &\href{http://www-personal.umich.edu/~mejn/netdata/}{url} \\ \hline 
$11$&{\tt College football} & $115$ & $613$ & $0.119$ & \cmark & \cmark & \cmark & \cite{girvan2002community} &\href{http://www-personal.umich.edu/~mejn/netdata/}{url} \\ \hline 
$12$&{\tt S 208$^*$} & $122$ & $189$ & $0.537$ & \cmark & \cmark & \cmark & \cite{milo2004superfamilies} &\href{http://wws.weizmann.ac.il/mcb/UriAlon/index.php?q=download/collection-complex-networks}{url} \\ \hline 
$13$&{\tt High school, 2011} & $126$ & $1,709$ & $0.033$ & \cmark & \cmark & \cmark & \cite{fournet2014contact} &\href{http://www.sociopatterns.org/datasets/high-school-dynamic-contact-networks/}{url} \\ \hline 
$14$&{\tt Bay Dry} & $128$ & $2,106$ & $0.027$ & \cmark & \cmark & \cmark & \cite{ulanowicz1998network, konect} &\href{http://konect.uni-koblenz.de/networks/foodweb-baydry}{url} \\ \hline 
$15$&{\tt Bay Wet} & $128$ & $2,075$ & $0.026$ & \cmark & \cmark & \cmark & \cite{konect} &\href{http://konect.uni-koblenz.de/networks/foodweb-baywet}{url} \\ \hline 
$16$&{\tt Radoslaw Email} & $167$ & $3,250$ & $0.017$ & \cmark & \cmark & \cmark & \cite{radoslaw, konect} &\href{http://konect.uni-koblenz.de/networks/radoslaw_email}{url} \\ \hline 
$17$&{\tt High school, 2012} & $180$ & $2,220$ & $0.039$ & \cmark & \cmark & \cmark & \cite{fournet2014contact} &\href{http://www.sociopatterns.org/datasets/high-school-dynamic-contact-networks/}{url} \\ \hline 
$18$&{\tt Little Rock Lake} & $183$ & $2,434$ & $0.028$ & \cmark & \cmark & \cmark & \cite{martinez1991artifacts, konect} &\href{http://konect.uni-koblenz.de/networks/maayan-foodweb}{url} \\ \hline 
$19$&{\tt Jazz} & $198$ & $2,742$ & $0.029$ & \cmark & \cmark & \cmark & \cite{gleiser2003community} &\href{http://deim.urv.cat/~alexandre.arenas/data/welcome.htm}{url} \\ \hline 
$20$&{\tt S 420$^*$} & $252$ & $399$ & $0.558$ & \cmark & \cmark & \cmark & \cite{milo2004superfamilies} &\href{http://wws.weizmann.ac.il/mcb/UriAlon/index.php?q=download/collection-complex-networks}{url} \\ \hline 
$21$&{\tt C. Elegans, neural} & $297$ & $2,148$ & $0.050$ & \cmark & \cmark & \cmark & \cite{watts1998collective} &\href{http://www-personal.umich.edu/~mejn/netdata/}{url} \\ \hline 
$22$&{\tt Network Science} & $379$ & $914$ & $0.247$ & \cmark & \cmark & \cmark & \cite{newman2006finding} &\href{http://www-personal.umich.edu/~mejn/netdata/}{url} \\ \hline 
$23$&{\tt Dublin} & $410$ & $2,765$ & $0.069$ & \cmark & \cmark & \cmark & \cite{isella2011s, konect} &\href{http://konect.uni-koblenz.de/networks/sociopatterns-infectious}{url} \\ \hline 
$24$&{\tt US Air Trasportation} & $500$ & $2,980$ & $0.027$ & \cmark & \cmark & \cmark & \cite{colizza2007reaction} &\href{https://sites.google.com/site/cxnets/usairtransportationnetwork}{url} \\ \hline 
$25$&{\tt S 838$^*$} & $512$ & $819$ & $0.564$ & \cmark & \cmark & \cmark & \cite{milo2004superfamilies} &\href{http://wws.weizmann.ac.il/mcb/UriAlon/index.php?q=download/collection-complex-networks}{url} \\ \hline 
$26$&{\tt Yeast, transcription} & $662$ & $1,062$ & $0.251$ & \cmark & \cmark & \cmark & \cite{milo2002network} &\href{http://wws.weizmann.ac.il/mcb/UriAlon/index.php?q=download/collection-complex-networks}{url} \\ \hline 
$27$&{\tt URV email} & $1,133$ & $5,451$ & $0.064$ & \cmark & \cmark & \cmark & \cite{guimera2003self} &\href{http://deim.urv.cat/~alexandre.arenas/data/welcome.htm}{url} \\ \hline 
$28$&{\tt Political blogs} & $1,222$ & $16,714$ & $0.016$ & \cmark & \cmark & \cmark & \cite{adamic2005political} &\href{http://www-personal.umich.edu/~mejn/netdata/}{url} \\ \hline 
$29$&{\tt Air traffic} & $1,226$ & $2,408$ & $0.201$ & \cmark & \cmark & \cmark & \cite{konect} &\href{http://konect.uni-koblenz.de/networks/maayan-faa}{url} \\ \hline 
$30$&{\tt Yeast, protein} & $1,458$ & $1,948$ & $0.341$ & \cmark & \cmark & \cmark & \cite{jeong2001lethality} &\href{http://www3.nd.edu/~networks/resources.htm}{url} \\ \hline 
$31$&{\tt Petster, hamster} & $1,788$ & $12,476$ & $0.027$ & \cmark & \cmark & \cmark & \cite{konect} &\href{http://konect.uni-koblenz.de/networks/petster-friendships-hamster}{url} \\ \hline 
$32$&{\tt UC Irvine} & $1,893$ & $13,835$ & $0.024$ & \cmark & \cmark & \cmark & \cite{opsahl2009clustering, konect} &\href{http://konect.uni-koblenz.de/networks/opsahl-ucsocial}{url} \\ \hline 
$33$&{\tt Yeast, protein} & $2,224$ & $6,609$ & $0.082$ & \cmark & \cmark & \cmark & \cite{bu2003topological} &\href{http://vlado.fmf.uni-lj.si/pub/networks/data/bio/Yeast/Yeast.htm}{url} \\ \hline 
$34$&{\tt Japanese} & $2,698$ & $7,995$ & $0.032$ & \cmark & \cmark & \cmark & \cite{milo2004superfamilies} &\href{http://wws.weizmann.ac.il/mcb/UriAlon/index.php?q=download/collection-complex-networks}{url} \\ \hline 
$35$&{\tt Open flights} & $2,905$ & $15,645$ & $0.021$ & \cmark & \cmark & \cmark & \cite{opsahl2010node, konect} &\href{http://konect.uni-koblenz.de/networks/opsahl-openflights}{url} \\ \hline 
$36$&{\tt GR-QC, 1993-2003} & $4,158$ & $13,422$ & $0.100$ & \cmark & \cmark & \cmark & \cite{leskovec2007graph} &\href{http://snap.stanford.edu/data/ca-GrQc.html}{url} \\ \hline 
$37$&{\tt Tennis} & $4,338$ & $81,865$ & $0.007$ & \cmark & \cmark & \cmark & \cite{radicchi2011best} &\href{-}{url} \\ \hline 
$38$&{\tt US Power grid$^*$} & $4,941$ & $6,594$ & $0.848$ & \cmark & \cmark & \cmark & \cite{watts1998collective} &\href{http://www-personal.umich.edu/~mejn/netdata/}{url} \\ \hline 
$39$&{\tt HT09} & $5,352$ & $18,481$ & $0.026$ & \cmark & \cmark & \cmark & \cite{isella2011s} &\href{http://www.sociopatterns.org/datasets/hypertext-2009-dynamic-contact-network/}{url} \\ \hline 
$40$&{\tt Hep-Th, 1995-1999} & $5,835$ & $13,815$ & $0.129$ & \cmark & \cmark & \cmark & \cite{newman2001structure} &\href{http://www-personal.umich.edu/~mejn/netdata/}{url} \\ \hline 
\end{tabular}
\end{center}
\caption{Summary table for real-world networks. The index appearing on the leftmost column serves only as a counter for the total number of networks analyzed. The following columns respectively report: the name of the network, the number of nodes in the giant component, the numer of edges in the giant component, the best estimate of the epidemic threshold, indications of whether the network has been included in the analysis for the subcritical, critical and supercritical regimes (\cmark indicates inclusion, \xmark indicates exclusion), reference(s) of the paper where the network has been first analyzed, and url of where the network data have been downloaded (to open the web page in your browser, just click on the word {\it url}).
Networks marked with $^*$ are spatially embedded networks.}
\label{table1}
\end{table*}

\clearpage
\begin{table*}
\begin{center}\begin{tabular}{|r|l|r|r|r|c|c|c|r|r|} \hline
$\#$ & network & $N$ & $E$ & $\lambda_c$ & Sub. & Cri. & Sup.  & Refs. & Url \\ 
 \hline 
$41$&{\tt Reactome} & $5,973$ & $145,778$ & $0.007$ & \cmark & \cmark & \cmark & \cite{joshi2005reactome, konect} &\href{http://konect.uni-koblenz.de/networks/reactome}{url} \\ \hline 
$42$&{\tt Jung} & $6,120$ & $50,290$ & $0.008$ & \cmark & \cmark & \cmark & \cite{vsubelj2012software, konect} &\href{http://konect.uni-koblenz.de/networks/subelj_jung-j}{url} \\ \hline 
$43$&{\tt Gnutella, Aug. 8, 2002} & $6,299$ & $20,776$ & $0.055$ & \cmark & \cmark & \cmark & \cite{ripeanu2002mapping, leskovec2007graph} &\href{http://snap.stanford.edu/data/p2p-Gnutella08.html}{url} \\ \hline 
$44$&{\tt JDK} & $6,434$ & $53,658$ & $0.008$ & \cmark & \cmark & \cmark & \cite{konect} &\href{http://konect.uni-koblenz.de/networks/subelj_jdk}{url} \\ \hline 
$45$&{\tt AS Oregon} & $6,474$ & $12,572$ & $0.037$ & \cmark & \cmark & \cmark & \cite{leskovec2005graphs} &\href{http://snap.stanford.edu/data/as.html}{url} \\ \hline 
$46$&{\tt English} & $7,377$ & $44,205$ & $0.011$ & \cmark & \cmark & \cmark & \cite{milo2004superfamilies} &\href{http://wws.weizmann.ac.il/mcb/UriAlon/index.php?q=download/collection-complex-networks}{url} \\ \hline 
$47$&{\tt Gnutella, Aug. 9, 2002} & $8,104$ & $26,008$ & $0.053$ & \cmark & \cmark & \cmark & \cite{ripeanu2002mapping, leskovec2007graph} &\href{http://snap.stanford.edu/data/p2p-Gnutella09.html}{url} \\ \hline 
$48$&{\tt French} & $8,308$ & $23,832$ & $0.023$ & \cmark & \cmark & \cmark & \cite{milo2004superfamilies} &\href{http://wws.weizmann.ac.il/mcb/UriAlon/index.php?q=download/collection-complex-networks}{url} \\ \hline 
$49$&{\tt Hep-Th, 1993-2003} & $8,638$ & $24,806$ & $0.087$ & \cmark & \cmark & \cmark & \cite{leskovec2007graph} &\href{http://snap.stanford.edu/data/ca-HepTh.html}{url} \\ \hline 
$50$&{\tt Gnutella, Aug. 6, 2002} & $8,717$ & $31,525$ & $0.071$ & \cmark & \cmark & \cmark & \cite{ripeanu2002mapping, leskovec2007graph} &\href{http://snap.stanford.edu/data/p2p-Gnutella06.html}{url} \\ \hline 
$51$&{\tt Gnutella, Aug. 5, 2002} & $8,842$ & $31,837$ & $0.066$ & \cmark & \cmark & \cmark & \cite{ripeanu2002mapping, leskovec2007graph} &\href{http://snap.stanford.edu/data/p2p-Gnutella05.html}{url} \\ \hline 
$52$&{\tt PGP} & $10,680$ & $24,316$ & $0.061$ & \cmark & \cmark & \cmark & \cite{boguna2004models} &\href{http://deim.urv.cat/~alexandre.arenas/data/welcome.htm}{url} \\ \hline 
$53$&{\tt Gnutella, August 4 2002} & $10,876$ & $39,994$ & $0.078$ & \cmark & \cmark & \cmark & \cite{ripeanu2002mapping, leskovec2007graph} &\href{http://snap.stanford.edu/data/p2p-Gnutella04.html}{url} \\ \hline 
$54$&{\tt Hep-Ph, 1993-2003} & $11,204$ & $117,619$ & $0.006$ & \cmark & \cmark & \cmark & \cite{leskovec2007graph} &\href{http://snap.stanford.edu/data/ca-HepPh.html}{url} \\ \hline 
$55$&{\tt Spanish} & $11,558$ & $43,050$ & $0.013$ & \cmark & \cmark & \cmark & \cite{milo2004superfamilies} &\href{http://wws.weizmann.ac.il/mcb/UriAlon/index.php?q=download/collection-complex-networks}{url} \\ \hline 
$56$&{\tt DBLP, citations} & $12,495$ & $49,563$ & $0.032$ & \cmark & \cmark & \cmark & \cite{ley2002dblp, konect} &\href{http://konect.uni-koblenz.de/networks/dblp-cite}{url} \\ \hline 
$57$&{\tt Email} & $12,625$ & $20,362$ & $0.031$ & \cmark & \cmark & \cmark & \cite{Kitsak10app} &\href{http://www-levich.engr.ccny.cuny.edu/webpage/hmakse/software-and-data/}{url} \\ \hline 
$58$&{\tt Spanish} & $12,643$ & $55,019$ & $0.011$ & \cmark & \cmark & \cmark & \cite{konect} &\href{http://konect.uni-koblenz.de/networks/lasagne-spanishbook}{url} \\ \hline 
$59$&{\tt Cond-Mat, 1995-1999} & $13,861$ & $44,619$ & $0.076$ & \cmark & \cmark & \cmark & \cite{newman2001structure} &\href{http://www-personal.umich.edu/~mejn/netdata/}{url} \\ \hline 
$60$&{\tt Astrophysics} & $14,845$ & $119,652$ & $0.019$ & \cmark & \cmark & \cmark & \cite{newman2001structure} &\href{http://www-personal.umich.edu/~mejn/netdata/}{url} \\ \hline 
$61$&{\tt Google} & $15,763$ & $148,585$ & $0.007$ & \cmark & \cmark & \cmark & \cite{palla2007directed} &\href{http://cfinder.org}{url} \\ \hline 
$62$&{\tt AstroPhys, 1993-2003} & $17,903$ & $196,972$ & $0.013$ & \cmark & \cmark & \cmark & \cite{leskovec2007graph} &\href{http://snap.stanford.edu/data/ca-AstroPh.html}{url} \\ \hline 
$63$&{\tt Cond-Mat, 1993-2003} & $21,363$ & $91,286$ & $0.040$ & \cmark & \cmark & \cmark & \cite{leskovec2007graph} &\href{http://snap.stanford.edu/data/ca-CondMat.html}{url} \\ \hline 
$64$&{\tt Gnutella, Aug. 25, 2002} & $22,663$ & $54,693$ & $0.122$ & \cmark & \cmark & \cmark & \cite{ripeanu2002mapping, leskovec2007graph} &\href{http://snap.stanford.edu/data/p2p-Gnutella25.html}{url} \\ \hline 
$65$&{\tt Internet} & $22,963$ & $48,436$ & $0.020$ & \cmark & \cmark & \cmark & - &\href{http://www-personal.umich.edu/~mejn/netdata/}{url} \\ \hline 
$66$&{\tt Thesaurus} & $23,132$ & $297,094$ & $0.011$ & \cmark & \cmark & \cmark & \cite{kiss1973associative, konect} &\href{http://konect.uni-koblenz.de/networks/eat}{url} \\ \hline 
$67$&{\tt Cora} & $23,166$ & $89,157$ & $0.050$ & \cmark & \cmark & \cmark & \cite{vsubelj2013model, konect} &\href{http://konect.uni-koblenz.de/networks/subelj_cora}{url} \\ \hline 
$68$&{\tt Linux, mailing list} & $24,567$ & $158,164$ & $0.006$ & \cmark & \cmark & \cmark & \cite{konect} &\href{http://konect.uni-koblenz.de/networks/lkml-reply}{url} \\ \hline 
$69$&{\tt AS Caida} & $26,475$ & $53,381$ & $0.021$ & \cmark & \cmark & \cmark & \cite{leskovec2005graphs} &\href{http://snap.stanford.edu/data/as-caida.html}{url} \\ \hline 
$70$&{\tt Gnutella, Aug. 24, 2002} & $26,498$ & $65,359$ & $0.105$ & \cmark & \cmark & \cmark & \cite{ripeanu2002mapping, leskovec2007graph} &\href{http://snap.stanford.edu/data/p2p-Gnutella24.html}{url} \\ \hline 
$71$&{\tt Hep-Th, citations} & $27,400$ & $352,021$ & $0.011$ & \cmark & \cmark & \cmark & \cite{leskovec2007graph, konect} &\href{http://konect.uni-koblenz.de/networks/cit-HepTh}{url} \\ \hline 
$72$&{\tt Cond-Mat, 1995-2003} & $27,519$ & $116,181$ & $0.039$ & \cmark & \cmark & \cmark & \cite{newman2001structure} &\href{http://www-personal.umich.edu/~mejn/netdata/}{url} \\ \hline 
$73$&{\tt Digg} & $29,652$ & $84,781$ & $0.041$ & \cmark & \cmark & \cmark & \cite{de2009social, konect} &\href{http://konect.uni-koblenz.de/networks/munmun_digg_reply}{url} \\ \hline 
$74$&{\tt Linux, soft.} & $30,817$ & $213,208$ & $0.007$ & \cmark & \cmark & \cmark & \cite{konect} &\href{http://konect.uni-koblenz.de/networks/linux}{url} \\ \hline 
$75$&{\tt Enron} & $33,696$ & $180,811$ & $0.011$ & \cmark & \cmark & \cmark & \cite{leskovec2009community} &\href{http://snap.stanford.edu/data/email-Enron.html}{url} \\ \hline 
$76$&{\tt Hep-Ph, citations} & $34,401$ & $420,784$ & $0.015$ & \cmark & \cmark & \cmark & \cite{leskovec2007graph, konect} &\href{http://konect.uni-koblenz.de/networks/cit-HepPh}{url} \\ \hline 
$77$&{\tt Cond-Mat, 1995-2005} & $36,458$ & $171,735$ & $0.027$ & \cmark & \cmark & \cmark & \cite{newman2001structure} &\href{http://www-personal.umich.edu/~mejn/netdata/}{url} \\ \hline 
$78$&{\tt Gnutella, Aug. 30, 2002} & $36,646$ & $88,303$ & $0.101$ & \cmark & \cmark & \cmark & \cite{ripeanu2002mapping, leskovec2007graph} &\href{http://snap.stanford.edu/data/p2p-Gnutella30.html}{url} \\ \hline 
$79$&{\tt Adult IMDB} & $47,719$ & $1,098,451$ & $0.002$ & \cmark & \cmark & \cmark & \cite{Kitsak10app} &\href{http://www-levich.engr.ccny.cuny.edu/webpage/hmakse/software-and-data/}{url} \\ \hline 
$80$&{\tt Slashdot} & $51,083$ & $116,573$ & $0.026$ & \cmark & \cmark & \cmark & \cite{gomez2008statistical, konect} &\href{http://konect.uni-koblenz.de/networks/slashdot-threads}{url} \\ \hline 
\end{tabular}
\end{center}
\caption{Summary table for real-world networks. The index appearing on the leftmost column serves only as a counter for the total number of networks analyzed. The following columns respectively report: the name of the network, the number of nodes in the giant component, the numer of edges in the giant component, the best estimate of the epidemic threshold, indications of whether the network has been included in the analysis for the subcritical, critical and supercritical regimes (\cmark indicates inclusion, \xmark indicates exclusion), reference(s) of the paper where the network has been first analyzed, and url of where the network data have been downloaded (to open the web page in your browser, just click on the word {\it url}).}
\label{table2}
\end{table*}

\clearpage
\begin{table*}
\begin{center}\begin{tabular}{|r|l|r|r|r|c|c|c|r|r|} \hline
$\#$ & network & $N$ & $E$ & $\lambda_c$ & Sub. & Cri. & Sup.  & Refs. & Url \\ 
 \hline 
$81$&{\tt Gnutella, Aug. 31, 2002} & $62,561$ & $147,878$ & $0.100$ & \cmark & \cmark & \cmark & \cite{ripeanu2002mapping, leskovec2007graph} &\href{http://snap.stanford.edu/data/p2p-Gnutella31.html}{url} \\ \hline 
$82$&{\tt Facebook} & $63,392$ & $816,886$ & $0.009$ & \cmark & \cmark & \cmark & \cite{viswanath2009evolution} &\href{http://socialnetworks.mpi-sws.org/data-wosn2009.html}{url} \\ \hline 
$83$&{\tt Epinions} & $75,877$ & $405,739$ & $0.007$ & \cmark & \cmark & \cmark & \cite{richardson2003trust, konect} &\href{http://konect.uni-koblenz.de/networks/soc-Epinions1}{url} \\ \hline 
$84$&{\tt Slashdot zoo} & $79,116$ & $467,731$ & $0.009$ & \cmark & \cmark & \cmark & \cite{kunegis2009slashdot, konect} &\href{http://konect.uni-koblenz.de/networks/slashdot-zoo}{url} \\ \hline 
$85$&{\tt Wikipedia, edits} & $113,123$ & $2,025,910$ & $0.003$ & \cmark & \cmark & \xmark & \cite{brandes2010structural, konect} &\href{http://konect.uni-koblenz.de/networks/wikiconflict}{url} \\ \hline 
$86$&{\tt Gowalla} & $196,591$ & $950,327$ & $0.008$ & \cmark & \cmark & \xmark & \cite{cho2011friendship, konect} &\href{http://konect.uni-koblenz.de/networks/loc-gowalla_edges}{url} \\ \hline 
$87$&{\tt EU email} & $224,832$ & $339,925$ & $0.013$ & \cmark & \cmark & \xmark & \cite{leskovec2007graph, konect} &\href{http://konect.uni-koblenz.de/networks/email-EuAll}{url} \\ \hline 
$88$&{\tt Amazon, Mar. 2, 2003} & $262,111$ & $899,792$ & $0.102$ & \cmark & \cmark & \xmark & \cite{leskovec2007dynamics} &\href{http://snap.stanford.edu/data/amazon0302.html}{url} \\ \hline 
$89$&{\tt DBLP, collaborations} & $317,080$ & $1,049,866$ & $0.037$ & \cmark & \cmark & \xmark & \cite{ley2002dblp, konect} &\href{http://konect.uni-koblenz.de/networks/dblp_coauthor}{url} \\ \hline 
$90$&{\tt Web Notre Dame} & $325,729$ & $1,090,108$ & $0.011$ & \cmark & \cmark & \xmark & \cite{albert1999internet} &\href{http://www3.nd.edu/~networks/resources.htm}{url} \\ \hline 
$91$&{\tt MathSciNet} & $332,689$ & $820,644$ & $0.048$ & \cmark & \cmark & \xmark & \cite{palla2008fundamental} &\href{http://cfinder.org}{url} \\ \hline 
$92$&{\tt CiteSeer} & $365,154$ & $1,721,981$ & $0.026$ & \cmark & \cmark & \xmark & \cite{bollacker1998citeseer, konect} &\href{http://konect.uni-koblenz.de/networks/citeseer}{url} \\ \hline 
%$93$&{\tt Actor coll. net.} & $374,511$ & $15,014,839$ & $0.002$ & \cmark & \xmark & \xmark & \cite{barabasi1999emergence, konect} &\href{http://konect.uni-koblenz.de/networks/actor-collaboration}{url} \\ \hline 
$93$&{\tt Amazon, Mar. 12, 2003} & $400,727$ & $2,349,869$ & $0.040$ & \cmark & \cmark & \xmark & \cite{leskovec2007dynamics} &\href{http://snap.stanford.edu/data/amazon0312.html}{url} \\ \hline 
$94$&{\tt Amazon, Jun. 6, 2003} & $403,364$ & $2,443,311$ & $0.036$ & \cmark & \cmark & \xmark & \cite{leskovec2007dynamics} &\href{http://snap.stanford.edu/data/amazon0601.html}{url} \\ \hline 
$95$&{\tt Amazon, May 5, 2003} & $410,236$ & $2,439,437$ & $0.038$ & \cmark & \cmark & \xmark & \cite{leskovec2007dynamics} &\href{http://snap.stanford.edu/data/amazon0505.html}{url} \\ \hline 
\end{tabular}
\end{center}
\caption{Summary table for real-world networks. The index appearing on the leftmost column serves only as a counter for the total number of networks analyzed. The following columns respectively report: the name of the network, the number of nodes in the giant component, the numer of edges in the giant component, the best estimate of the epidemic threshold, indications of whether the network has been included in the analysis for the subcritical, critical and supercritical regimes (\cmark indicates inclusion, \xmark indicates exclusion), reference(s) of the paper where the network has been first analyzed, and url of where the network data have been downloaded (to open the web page in your browser, just click on the word {\it url}).}
\label{table3}
\end{table*}

\bibliography{SM_biblio}

%%%%%%%%%%%%%%%%%%%%%


\end{document}